\documentclass[twoside,twocolumn,9pt]{article}
\usepackage{extsizes}
\usepackage[super,sort&compress,comma]{natbib} 
\usepackage[version=3]{mhchem}
\usepackage[left=1.5cm, right=1.5cm, top=1.785cm, bottom=2.0cm]{geometry}
\usepackage{balance}
\usepackage{mathptmx}
\usepackage{sectsty}
\usepackage{graphicx} 
\usepackage{lastpage}
\usepackage[format=plain,justification=justified,singlelinecheck=false,font={stretch=1.125,small,sf},labelfont=bf,labelsep=space]{caption}
\usepackage{float}
\usepackage{fancyhdr}
\usepackage{fnpos}
\usepackage[english]{babel}
\addto{\captionsenglish}{%
  
}
\usepackage{array}
\usepackage{droidsans}
\usepackage{charter}
\usepackage[T1]{fontenc}
\usepackage[usenames,dvipsnames]{xcolor}
\usepackage{setspace}
\usepackage[compact]{titlesec}
\usepackage{hyperref}
\usepackage{multirow}
\usepackage{makecell}

\usepackage{epstopdf}

\definecolor{cream}{RGB}{222,217,201}

\begin{document}

\pagestyle{fancy}
\thispagestyle{plain}
\fancypagestyle{plain}{
\renewcommand{\headrulewidth}{0pt}
}

\makeFNbottom
\makeatletter
\renewcommand\LARGE{\@setfontsize\LARGE{15pt}{17}}
\renewcommand\Large{\@setfontsize\Large{12pt}{14}}
\renewcommand\large{\@setfontsize\large{10pt}{12}}
\renewcommand\footnotesize{\@setfontsize\footnotesize{7pt}{10}}
\renewcommand\scriptsize{\@setfontsize\scriptsize{7pt}{7}}
\makeatother

\renewcommand{\thefootnote}{\fnsymbol{footnote}}
\renewcommand\footnoterule{\vspace*{1pt}%
\color{cream}\hrule width 3.5in height 0.4pt \color{black} \vspace*{5pt}} 
\setcounter{secnumdepth}{5}

\makeatletter 
\renewcommand\@biblabel[1]{#1}            
\renewcommand\@makefntext[1]%
{\noindent\makebox[0pt][r]{\@thefnmark\,}#1}
\makeatother 
\renewcommand{\figurename}{\small{Fig.}~}
\sectionfont{\sffamily\Large}
\subsectionfont{\normalsize}
\subsubsectionfont{\bf}
\setstretch{1.125} 
\setlength{\skip\footins}{0.8cm}
\setlength{\footnotesep}{0.25cm}
\setlength{\jot}{10pt}
\titlespacing*{\section}{0pt}{4pt}{4pt}
\titlespacing*{\subsection}{0pt}{15pt}{1pt}

\fancyfoot{}
\fancyfoot[LO,RE]{\vspace{-7.1pt}\includegraphics[height=9pt]{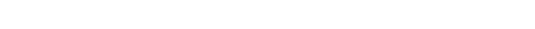}}
\fancyfoot[CO]{\vspace{-7.1pt}\hspace{13.2cm}\includegraphics{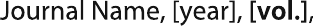}}
\fancyfoot[CE]{\vspace{-7.2pt}\hspace{-14.2cm}\includegraphics{head_foot/RF}}
\fancyfoot[RO]{\footnotesize{\sffamily{1--\pageref{LastPage} ~\textbar  \hspace{2pt}\thepage}}}
\fancyfoot[LE]{\footnotesize{\sffamily{\thepage~\textbar\hspace{3.45cm} 1--\pageref{LastPage}}}}
\fancyhead{}
\renewcommand{\headrulewidth}{0pt} 
\renewcommand{\footrulewidth}{0pt}
\setlength{\arrayrulewidth}{1pt}
\setlength{\columnsep}{6.5mm}
\setlength\bibsep{1pt}

\makeatletter 
\newlength{\figrulesep} 
\setlength{\figrulesep}{0.5\textfloatsep} 

\newcommand{\topfigrule}{\vspace*{-1pt}%
\noindent{\color{cream}\rule[-\figrulesep]{\columnwidth}{1.5pt}} }

\newcommand{\botfigrule}{\vspace*{-2pt}%
\noindent{\color{cream}\rule[\figrulesep]{\columnwidth}{1.5pt}} }

\newcommand{\dblfigrule}{\vspace*{-1pt}%
\noindent{\color{cream}\rule[-\figrulesep]{\textwidth}{1.5pt}} }

\makeatother

\twocolumn[
  \begin{@twocolumnfalse}
\vspace{1em}
\sffamily
\begin{tabular}{m{4.5cm} p{13.5cm} }

& \noindent\LARGE{\textbf{Computationally driven discovery of SARS-CoV-2 M$^{pro}$ inhibitors: from design to experimental validation$^\dag$}} \\

 & \vspace{0.3cm} \\

 & \noindent\large {Léa El Khoury,\textit{$^{a}$}\textit{$^{\ddag}$}
 Zhifeng Jing,\textit{$^{a}$}\textit{$^{\ddag}$}
 Alberto Cuzzolin,\textit{$^{b}$} 
 Alessandro Deplano,\textit{$^{c}$}
 Daniele Loco,\textit{$^{a}$}
 Boris Sattarov,\textit{$^{a}$}
 Florent Hédin,\textit{$^{a}$}
 Sebastian Wendeborn,\textit{$^{d}$}
 Chris Ho,\textit{$^{a}$}
 Dina El Ahdab,\textit{$^{o}$}
 Theo Jaffrelot Inizan,\textit{$^{o}$}
 Mattia Sturlese,\textit{$^{g}$}
 Alice Sosic,\textit{$^{e}$}
Martina Volpiana,\textit{$^{e}$}
Angela Lugato,\textit{$^{e}$}
Marco Barone,\textit{$^{e}$}
Barbara Gatto,\textit{$^{e}$}
Maria Ludovica Macchia,\textit{$^{e}$}
Massimo Bellanda ,\textit{$^{f}$}
Roberto Battistutta,\textit{$^{f}$}
Cristiano Salata,\textit{$^{h}$}
Ivan Kondratov,\textit{$^{i}$}
Rustam Iminov,\textit{$^{i}$}
Andrii Khairulin,\textit{$^{i}$}
Yaroslav Mykhalonok,\textit{$^{i}$}
Anton Pochepko,\textit{$^{i}$}
Volodymyr Chashka-Ratushnyi,\textit{$^{i}$}
Iaroslava Kos,\textit{$^{i}$}
Stefano Moro,\textit{$^{g}$}
Matthieu Montes,\textit{$^{l}$}
Pengyu Ren,\textit{$^{m}$}
Jay W. Ponder,\textit{$^{n,o}$}
Louis Lagardère,\textit{$^{p}$}
Jean-Philip Piquemal$^{\ast}$\textit{$^{p,q}$} and Davide  Sabbadin$^{\ast}$\textit{$^{a,d}$}} \\

\end{tabular}

 \end{@twocolumnfalse} \vspace{0.6cm}

  ]

\renewcommand*\rmdefault{bch}\normalfont\upshape
\rmfamily
\section*{}
\vspace{-1cm}


\footnotetext{\textit{$^{a}$~Qubit Pharmaceuticals, Incubateur Paris Biotech Santé, 24 rue du Faubourg Saint Jacques
75014 Paris, France;  Contact email: davide@qubit-pharmaceuticals.com (DS)}}

\footnotetext{\textit{$^{b}$Current affiliation: Chiesi farmaceutici S.p.A, Nuovo centro ricerche, Largo belloli 11a, 43122, Parma (Italy)}}

\footnotetext{\textit{$^{c}$Current affiliation: Pharmacelera, Torre R, 4a planta, Despatx A05, Parc Cientific de Barcelona, Baldiri Reixac 8, 08028 Barcelona, (Spain)}}

\footnotetext{\textit{$^{d}$University of Applied Sciences and Arts Northwestern Switzerland - School of LifeSciences, Hofackerstrasse 30, CH-4132 Muttenz, Switzerland }}

\footnotetext{\textit{$^{e}$Department of Pharmaceutical and Pharmacological Sciences, University of Padova, via Marzolo 5, 35131, Padova (Italy)}}

\footnotetext{\textit{$^{f}$Department of Chemistry, University of Padova, via Marzolo 1, 35131, Padova (Italy)}}

\footnotetext{\textit{$^{g}$Molecular Modeling Section, Department of Pharmaceutical and Pharmacological Sciences, University of Padua, via F. Marzolo 5, 35131, Padova, (Italy)}}

\footnotetext{\textit{$^{h}$Department of Molecular Medicine, University of Padua, via Gabelli 63, 35121, Padova, (Italy)}}

\footnotetext{\textit{$^{i}$Enamine LTD, 78 Chervonotkats‘ka str., Kyiv 02094 (Ukraine)}}

\footnotetext{\textit{$^{l}$Laboratoire GBCM, EA7528, Conservatoire National des Arts et Métiers, Hesam Université, 2 Rue Conte, 75003 Paris, France}}

\footnotetext{\textit{$^{m}$University of Texas at Austin, Department of Biomedical Engineering, TX 78712, USA}}

\footnotetext{\textit{$^{n}$Department of Chemistry, Washington University in Saint Louis, MO 63130, USA}}

\footnotetext{\textit{$^{o}$Department of Biochemistry and Molecular Biophysics, Washington University School of Medicine, MO 63110, USA}}

\footnotetext{\textit{$^{p}$Sorbonne Université, Laboratoire de Chimie Théorique, UMR 7616 CNRS, 75005, Paris,France;Contact email: jean-philip.piquemal@sorbonne-universite.fr (JPP)}}

\footnotetext{\textit{$^{q}$Institut Universitaire de France, 75005, Paris,France.}}

\footnotetext{\textit{$^{\ddag}$These authors contributed equally to this work}}
\footnotetext{\dag~Electronic Supplementary Information (ESI) available: [details of any supplementary information available should be included here]. See DOI: 00.0000/00000000.}



\sffamily{\textbf{We report a fast-track computationally-driven discovery of new SARS-CoV2 Main Protease (M$^{pro}$) inhibitors whose potency range from mM for initial non-covalent ligands to sub-$\mu${M} for the final covalent compound (IC50=830 ± 50 nM). The project extensively relied on high-resolution all-atom molecular dynamics simulations and absolute binding free energy calculations performed using the polarizable AMOEBA force field. The study is complemented by extensive adaptive sampling simulations that are used to rationalize the different ligands binding poses through the explicit reconstruction of the ligand-protein conformation spaces. Machine Learning predictions are also performed to predict selected compound properties. While simulations extensively use High Performance Computing to strongly reduce time-to-solution, they were systematically coupled to Nuclear Magnetic Resonance experiments to drive synthesis and to \textit{in vitro} characterization of compounds. Such study highlights the power of \textit{in silico} strategies that rely on structure-based approaches for drug design and allows to address the protein conformational multiplicity problem. The proposed fluorinated tetrahydroquinolines open routes for further optimization of M$^{pro}$ inhibitors towards low nM affinities. \\ } }


\rmfamily 

\section*{Introduction}
Since December 2019, the Covid-19 global pandemic has put the entire world on edge.\cite{Zhou2020pneumonia,Wu2020new} The disease is due to a coronavirus (CoV) called SARS-CoV-2 (severe acute respiratory disease, SARS) that has triggered the start of an unprecedented research effort.\cite{D0CS01065K,von2021white,anie.202016961} While the vaccination strategy \cite{krammer2020sars} has been particularly successful with the rise of mRNA techniques, additional programs have been launched to obtain antivirals able to reduce the impact of COVID19 on ill patients. Despite these efforts, few potential treatments are presently available at the exception of the Paxlovid, a nirmatrelvir/ritonavir combo proposed by Pfizer.\cite{ledford2021covid} Due to the persistence of the pandemic, it remains essential to propose new antiviral drugs. A possible strategy consists in designing small molecules to interact with one of the main proteins of SARS-Cov2 virus, thus blocking its activity. Among the potential targets, the main protease protein, denoted as M$^{pro}$ or 3CLpro, is a primary choice \cite{zhang-2020} as it has no human homolog and it is well conserved among coronaviruses,\cite{yang2005design} especially in terms of the structure of its active site, catalytic dyad, and dimer interface. Furthermore, M$^{pro}$ is required to release viral proteins for particle assembly, and is thus essential to the virus replication cycle. 

Developing a new drug targeting the viral M$^{pro}$ is challenging as it requires extensive resources and the rate success is notoriously low.\cite{shermannreview} Relying on \textit{in silico} driven rational design could accelerate the process. In fact, it diminishes the cost by reducing the need for synthetic iterations while also providing an interpretation of the interactions occurring between the target protein and potential inhibitors.

It is important to note that theoretical modeling of M$^{pro}$ is challenging as the protein exhibits a high structural flexibility \cite{kneller2020structural,D1SC00145K,acs.jpclett.1c01460} leading to a high conformational complexity. M$^{pro}$ is also involved in a variety of complex protein-ligand-solvent interaction networks.\cite{D1SC00145K,acs.jpclett.1c01460} These challenges can be tackled using a high-resolution modeling approach \cite{D1SC00145K,acs.jpclett.1c01460} going beyond rigid docking procedures (see reference \cite{majordocking} for a detailed discussion of the difficulties of docking approaches in predicting the native binding modes of small molecules within M$^{pro}$ ).  

Many studies have been devoted to the design of new M$^{pro}$ inhibitors \cite{Li27381,D1SC03628A,D1CC00050K,anie.202016961}
\cite{D0CS01065K,shcherbakov_design_2021,glaab_pharmacophore_2021,jin_structure_2020,gossen_blueprint_2021,manandhar_discovery_2021,amendola_lead_2021,ghahremanpour_identification_2020,zhang_optimization_2021}
through joint computational and experimental approaches. In particular, a recent study by the Jorgensen group highlighted the usefulness of relative binding free energy computations (RBFE) as part of the drug design process.\cite{Jorgensen}

In this paper, we present a computationally-driven discovery and binding mode rationalization of new SARS-CoV-2 M$^{pro}$ inhibitors. In doing so, we build on our previous high-resolution M$^{pro}$ molecular dynamics studies.\cite{D1SC00145K,acs.jpclett.1c01460} Here, we explore more deeply some specific subpockets of the substrate binding site of the protease using absolute binding free energy (ABFE) calculations and adaptive sampling grounded on extensive molecular dynamics simulations with high-resolution polarizable force fields (PFF). Using the GPU-accelerated module \cite{adjoua2021tinker} (GPU=Graphics Processing Unit) of the Tinker-HP molecular dynamics package \cite{lagardere-2018} coupled to the AMOEBA PFF,\cite{ren2003polarizable,ponder2010current,shi2013polarizable,AMOEBAnuc} it has been shown that simulations can reach the required level of accuracy and $\mu$s timescales needed to explore 
the structural rearrangement and interactions profile of this flexible protein.\cite{D1SC00145K,acs.jpclett.1c01460} More precisely, the modeling of M$^{pro}$ necessitates the ability to evaluate at high-resolution various types of key interactions including hydrogen bonds, salt bridges, $\pi$ –$\pi$ stacking, and specific solvation effects. Long timescales are required to achieve sufficient sampling. This is now possible by using the large number of graphics processing units (GPUs) that are presently available on supercomputers and high-performance cloud computing platforms. In this study, we combine our computationally-driven strategy, using absolute binding free energy computations \cite{Jiao6290,doi:10.1021/jp210265d,doi:10.1021/acs.jpcb.8b03194,D1SC01887F,shi2021amoeba} and unsupervised adaptive sampling,\cite{D1SC00145K,acs.jpclett.1c01460} with machine learning-assisted property predictions, while providing extensive characterization experiments including nuclear magnetic resonance (NMR), mass spectrometry (MS), and FRET-based assays to evaluate the activity of the newly designed compounds. 

In the following, we introduce our design strategy, which led to non-covalent and covalent inhibitors of M$^{pro}$ (SI-Figure 1). Then, we describe how an interplay between experiments and molecular simulations allowed the discovery of a final compound (QUB-00006-Int-07) with a high affinity to the protease (IC50=830 ± 50 nM).

\section*{Computational details}
\subsection*{A. Systems preparation}
The protease dimer structure (PDB code: 7L11) was used for all the MD simulations and it was prepared at physiological pH (pH=7). This structure has a higher resolution (1.80 Å) than the PDB structure (PDB code: 6LU7) used in our previous work \cite{D1SC00145K} (resolution of 2.16 Å). Both structures are of the holo state in complex with covalent inhibitors, and the rotamers of the key residues at the catalytic site (Cys145, His41, His162, His163, His172) are virtually identical. The protonation states of His residues were assigned based on previous work \cite{Arafet2021}, where His41 and His80 are protonated at the delta carbon atom and all other His residues are epsilon-protonated, which is favorable for the substrate binding\cite{Arafet2021}. This is different from our previous work where His64 and His80 are protonated at the delta carbon atom and all other histidines are epsilon-protonated.\cite{D1SC00145K} All water molecules were retained except for those that might collide with the ligands. 
\subsection*{B. Simulations protocols}
All-atom simulations were performed using Qubit Pharmaceuticals' Atlas platform which enables the use of any type of High-Performance Computing (HPC) systems including cloud supercomputing infrastructures. Among its possibilities, Atlas has the ability to efficiently handle polarizable force field molecular dynamics simulations using a custom version of the multi-GPU module \cite{adjoua2021tinker}of the Tinker-HP molecular dynamics package \cite{lagardere-2018,Jolly_Duran_2019}, to perform docking runs using either Autodock-Vina \cite{Trott2010} or Autodock-GPU \cite{autodockgpu_2021}, and to enable machine learning predictions of molecular properties.
\subsubsection*{B.1. Molecular Dynamics simulations}
  All Tinker-HP MD simulations (for a total of several $\mu$s) were performed in mixed precision to benefit from a strong acceleration of simulations using GPUs.\cite{adjoua2021tinker} The AMOEBA polarizable force field \cite{ren2003polarizable, ponder2010current, shi2013polarizable, AMOEBAnuc} was used to describe the full systems including the protein, ions and water. Several utilities (TinkerTools) from Tinker 8 \cite{tinker8} were used. Periodic boundary conditions were applied within the framework of smooth particle mesh Ewald summation \cite{SPME,SPMEHP} with a grid of dimensions 120×120×120 using a cubic box with side lengths of 97 Å. The Ewald cutoff was set to 7 Å, and the van der Waals cutoff was 12 Å. Langevin molecular dynamics simulations were performed using the recently introduced BAOAB-RESPA1 integrator (10 fs outer timestep)\cite{baoabrespa1}, a preconditioned conjugate gradient polarization solver (with a 10$^{-5}$ convergence threshold) to solve polarization at each time step\cite{lagardere2015scalable}, hydrogen-mass repartitioning (HMR) and random initial velocities.  Absolute free energy simulations following a protocol described in the next section were performed as well as adaptive sampling runs that are also described further in the text. Absolute free energy computations  were both performed on the HPE Jean Zay Supercomputer (IDRIS, GENCI, France) and on Amazon Web Services (AWS). All adaptive sampling computations were performed using AWS. Simulations at AWS used both p3.2x (NVIDIA V100 GPU cards) and p4d.24xlarge (NVIDIA A100 GPU cards) instances whereas computations on the Jean Zay supercomputer were powered by V100 cards. 
\subsubsection*{B.2. Molecular Docking protocol}
The protonation states of the ligands 
were calculated at a neutral pH and the hydrogen atoms were added using Chimera. Next, we docked the ligands QUB-00006-Int-01(R) and QUB-00006-Int-01(S) into the M$^{pro}$ dimer structure using Autodock Vina 1.1.2 \cite{Trott2010}. AutoDock Vina requires pdbqt format for the input files of the receptor and the ligand. Therefore, using the scripts `prepare\_receptor4.py` (v 1.13) and `prepare\_ligand4.py` (v 1.10) provided by Autodock Tools \cite{Morris2009}, we generated pdbqt files corresponding to the receptor and the ligands, respectively. We set the exhaustiveness search to 100 and the num\_mode option to 50.\\
Since molecular docking could suggest reasonable potential binding modes, but does not always rank the most likely binding mode as the best docked pose \cite{Warren2006,majordocking}, we visually inspected the generated docked poses and chose an ensemble of binding poses with different binding orientations that we used to run MD and ABFE calculations in order to explore the binding mode of QUB-00006-Int-01, as described in the Results and discussion section.

\subsection*{B.3. Equilibration}
A detailed description of the equilibration protocol used for MD simulations can be found in SI.
\subsection*{B.4. High-resolution Adaptive Molecular dynamics simulations}
Starting from several binding poses as described above  
we ran adaptive sampling simulations using the AMOEBA force field \cite{ren2003polarizable,ponder2010current,shi2013polarizable,AMOEBAnuc} in order to explore their stability and more generally to explore the conformational space of the ligands in the pocket of the M$^{pro}$. Because of the flexibility of the pocket and the role it may play in the exploration of the potential binding modes of the ligand, we chose to keep the whole system (ligand+protein) flexible during this sampling phase. The restart strategy (similar to the one introduced in \cite{D1SC00145K}) was the following: first, all the previously generated conformations of the protein were loaded and aligned with MDTraj \cite{mcgibbon2015mdtraj}, then PCAs of the conformations of the ligand were computed using Scikitlearn \cite{pedregosa2011scikit} and these frames were projected on the first four PCAs. Finally, the same scheme as the one described in \cite{D1SC00145K} was used to generate new starting points, favoring points that were less explored during the previous phases. In practice, a first set of 5 simulations of 10 nanoseconds were done using different random seeds, then 4 iterations of 10 times 10 nanoseconds were generated using the adaptive sampling protocol described above, for a total of 450 nanoseconds.
\subsection*{B.5. Absolute binding free energy calculations}
In order to benefit from the high-accuracy evaluation of free energies using the AMOEBA force field,\cite{Jiao6290,doi:10.1021/jp210265d,doi:10.1021/acs.jpcb.8b03194,D1SC01887F,shi2021amoeba}, we used the same clustering algorithms as described above to analyze the adaptive molecular dynamics simulations. The largest clusters were used for absolute free energy calculations. The double-decoupling protocol and the Bennett acceptance ratio (BAR) \cite{BENNETT1976245} method were used to calculate the standard binding free energy for each binding pose.\cite{Jiao6290,shi2021amoeba} There were 27 or 26 thermodynamic states for the decoupling in complex phase or the aqueous phase. A distance restraint between two groups of atoms in the ligand and in the protein binding pocket was applied when decoupling the ligand in complex to accelerate the convergence when the ligand is fully decoupled, and the restraint was removed at an additional step at the full interaction state. A harmonic restraint with force constant 15.0 kcal/mol/Å$^2$ and radius 2.0 Å was used. An analytic correction was added to the binding free energy to account for the standard state at 1.0 mol/L in the fully decoupled state.  10 ns simulations were performed for each thermodynamic state for the simulations of M$^{pro}$ in complex with x0195, QUB-00006 (S), QUB-00006 (R), and QUB-00006-Int-07. For the simulations of M$^{pro}$ in complex with QUB-00006-Int-01 (R) and QUB-00006-Int-01 (S), we ran each thermodynamic state for 20 ns. We used the BAOAB-RESPA1 integrator with 10 fs timestep and we calculated the electrostatic interactions using Ewald summation with a real space cutoff of 7 Å. Van der Waals interactions were calculated using a cutoff of 12 Å with long-range correction.

\subsection*{C. Quantitative Structure-Property Relationship (QSPR) modeling: predicting solubility using machine learning}
Qubit Pharmaceuticals' Atlas internal machine learning-based QSPR module was used to predict water solubility (logS Molar) and octanol/water partition coefficient (logP). To build a water solubility QSPR predictor, AqSolDB dataset \cite{RN629} was used as a training set. To predict octanol/water partition coefficients (logP), the dataset from EPA’s OPERA\cite{mansouri_opera_2018} was used as a training set.

Selected datasets were preprocessed and standardized to some extent by authors of the corresponding publications. However, the need for additional processing was identified when doing exploratory data analysis. We discarded compounds with less than two carbon atoms and kept molecules with molecular weight between 50 and 750 daltons. Additional rules of fragments standardization developed at Qubit Pharmaceuticals were applied. 

\textbf{Similarity analysis.} 
Tanimoto similarity \cite{tanimoto} to the x0195 compound was calculated for each molecule using 
the MAACS fingerprint from the RDKit Open-Source Cheminformatics Software (https://www.rdkit.org). 
The Morgan Circular Fingerprint \cite{morgan} with radius=2 and nBits=2048 from RDKit was also tested and the results (not shown) have similar ranking of the compounds.

\section*{Experimental protocol}

\subsection*{A. Recombinant Expression of SARS-CoV-2 M$^{pro}$ in \emph{E. coli}}
The plasmid pGEX-6P-1 encoding SARS-CoV-2 M$^{pro}$ \cite{zhang-co} was a generous gift from Prof. Rolf Hilgenfeld, University of Lübeck, Lübeck, Germany. Protein expression and purification were adapted from Zhang, et al. \cite{zhang-co} The expression plasmid was transformed into \emph{E. coli} strain BL21 (DE3) and then pre-cultured in YT medium at 37 °C (100 $\mu$g/mL ampicillin) overnight. The pre-culture was used to inoculate fresh YT medium supplied with antibiotic and the cells were grown at 37 °C to an OD600 of 0.6–0.8 before induction of overexpression with 0.5 mM isopropyl-D-thiogalactoside (IPTG). After 5 h at 37 °C, cells were harvested by centrifugation (5000 g, 4 °C, 15 min) and frozen. The pellets were resuspended in buffer A (20 mM Tris, 150 mM NaCl, pH 7.8) supplemented with lysozyme, DNase I and PMSF for the lysis. The lysate was clarified by centrifugation at 12000 g at 4 °C for 1 h and loaded onto a HisTrap HP column (GE Healthcare) equilibrated with 98\% buffer A/2\% buffer B (20 mM Tris, 150 mM NaCl, 500 mM imidazole, pH 7.8). The column was washed with 95\% buffer A/5\% buffer B and then His-tagged M$^{pro}$ was eluted with a linear gradient of imidazole ranging from 25 mM to 500 mM. Pooled fractions containing target protein were subjected to buffer exchange with buffer A using a HiPrep 26/10 desalting column (GE Healthcare). Next, PreScission protease was added to remove the C-terminal His tag (20 $\mu$g of PreScission protease per mg of target protein) at 12 °C overnight. Protein solution was loaded onto a HisTrap HP column connected to a GSTtrap FF column (GE Healthcare) equilibrated in buffer A to remove the GST-tagged PreScission protease, the His-tag, and the uncleaved protein. M$^{pro}$ was finally purified with a Superdex 75 prep-grade 16/60 (GE Healthcare) SEC column equilibrated with buffer C (20 mM Tris, 150 mM NaCl, 1 mM EDTA, 1 mM DTT, pH 7.8). Fractions containing the target protein at high purity were pooled, concentrated at 25 mg/ml and flash-frozen in liquid nitrogen for storage in small aliquots at -80°C.

\subsection*{B. Protein characterization and enzymatic activity}
The molecular mass of the recombinant SARS-CoV-2 M$^{pro}$ was determined by direct infusion electrospray ionization mass spectrometry (ESI-MS) on a Xevo G2-XS QTOF mass spectrometer (Waters). Samples were diluted in 50\% acetonitrile with 0.1\% of formic acid to achieve a final 1 $\mu$M concentration of protein. The detected species displayed a mass of 33796.64 Da, which matches very closely the value of 33796.81 Da calculated from the theoretical full-length protein sequence (residues 1-306). To characterize the enzymatic activity of our recombinant M$^{pro}$, we adopted a FRET-based assay using the fluorogenic substrate 5-FAM-AVLQ\`~SGFRK(DABCYL)K (Proteogenix) harbouring the cleavage site of SARS-CoV-2 M$^{pro}$ (\`~ indicates the cleavage site). The fluorescence of the intact peptide is very low since the fluorophore 5-FAM and the quencher Dabcyl are in close proximity. When the substrate is cleaved by the protease, the fluorophore and the quencher are separated, increasing the fluorescence signal. Freshly unfrozen recombinant SARS-CoV-2 M$^{pro}$ was used in our assays. The assay was performed by mixing 0.05 $\mu$M M$^{pro}$ with different concentrations of substrate (1-128 $\mu$M) in the reaction buffer (20 mM Tris-HCl, 100 mM NaCl, 1 mM EDTA and 1 mM DTT, pH 7.3) in the final volume of 100 $\mu$L. Fluorescence intensity (Ex = 485 nm/Em = 535 nm) was monitored at 37 °C with a microplate reader VictorIII (Perkin Elmer) for 50 min. A calibration curve was created by measuring multiple concentrations (from 0.001 to 5 $\mu$M) of free fluorescein in a final volume of 100 $\mu$L reaction buffer. Initial velocities were determined from the linear section of the curve, and the corresponding relative fluorescence units per unit of time ($\Delta$RFU/s) was converted to the amount of the cleaved substrate per unit of time ($\mu$M/s) by fitting to the calibration curve of free fluorescein. Inner-filter effect corrections were applied for the kinetic measurements according to \cite{Liu-1999}. The catalytic efficiency kcat/km resulted in 4819± 399 s-1M-1, in line with literature data \cite{boceprevir_2020, zhang-co}.

\subsection*{C. Nuclear Magnetic Resonance}

All the NMR screening experiments were acquired with a Bruker Neo 600MHz spectrometer, equipped with nitrogen cooled Prodigy CryoProbe 5mm at 298K. The ligand binding was monitored by WaterLOGSY (wLogsy) \cite{dalvit-2001} and Saturation Transfer Difference (STD) \cite{std}   experiments in the presence and in the absence of the protein. Samples contained 10 µM of M$^{pro}$ and ligand concentration varying from100 µM to 2 mM of ligand dissolved in 150 mM NaCl, 20 mM Phosphate, 5\% D2O, 4\% DMSO-d6 (pH=7.3).
Water-LOGSY experiments were performed with a 180° inversion pulse applied to the water signal at 4.7ppm using a Gaussian-shaped selective pulse of 5ms. Each Water-LOGSY spectrum was acquired with 320 scans a mixing time of 1.5s and a relaxation delay of 4.5 s. STD experiments were acquired with 256 scans. Selective saturation of the protein at 0.4ppm frequency was carried out by a 2s pulse train (60 Gaussian pulses of 50ms separated by 1ms intervals) included in the relaxation delay and a 30ms spin-lock was used to reduce the broad background,protein signal. The estimation of the KD was achieved by a STD titration according to previously reported procedure and fitting the curves using OriginPro 2018 (Origin(Pro), Version 2018
by OriginLab Corporation, Northampton, MA, USA). The water suppression was achieved by the excitation sculpting pulse scheme.

\subsection*{D. Screening of potential M$^{pro}$ inhibitors and hits validation}
The FRET-based assay employed to test the enzymatic activity of the recombinant SARS-CoV-2 M$^{pro}$ was used to evaluate the ability of the compounds to inhibit its activity in vitro. In fact, inhibition of M$^{pro}$ by the tested compounds results in a decrease of the fluorescence signal compared to the M$^{pro}$ activity in the absence of an inhibitor. A preliminary screening was first performed at a single compound concentration to rapidly identify the compounds ability to inhibit M$^{pro}$ activity and to rank them according to their inhibitory activity. The protein was diluted in the reaction buffer (20 mM Tris-HCl, 100 mM NaCl, 1 mM EDTA and 1 mM DTT, pH 7.3) and pipetted into a 96-well plate to the final protein concentration of 0.02 µM in a final volume of 100 µL. Each compound at the final concentration of 100 µM was incubated with M$^{pro}$ for 20 minutes at room temperature.  After incubation, the peptide substrate (5 µM final) was added to initiate the reaction which was monitored for 50 min at 37°C. The final DMSO amount was 3,75\%. Two controls were prepared for each experiment: the peptide substrate in the absence of M$^{pro}$ (0\% M$^{pro}$ activity, hence minimal fluorescence intensity detected) and the reaction mixture in the absence of compound (100\% M$^{pro}$ activity, therefore maximal fluorescence intensity detected). Following the preliminary screening, the most active compounds (hits) were tested at increasing concentrations (0.25, 0.5, 1, 5, 25, 50, 100, 150 µM) to determine the dose-response curves and calculate IC50 values fitted by using GraphPad Prism 5 software. Each experiment was performed in triplicate and the results were used to calculate an average and a standard deviation.

\subsection*{E. Binding studies by Mass Spectrometry}
Samples were prepared by mixing appropriate volumes of M$^{pro}$ (10 µM final) with each compound in the reaction buffer (20 mM Tris-HCl, 100 mM NaCl, 1 mM EDTA and 1 mM DTT, pH 7.3). The final mixtures contained 1:1 or 10:1 of compound:protein molar ratio. Samples were incubated at room temperature for 20 min before analysis. Control experiments were performed on 10 µM solutions of M$^{pro}$ in the absence of compound. Mass spectrometric analyses were carried out in positive ion mode by ESI-MS under denaturing conditions i.e. water/acetonitrile 50:50 with 0.1\% formic acid on a Q-Tof Xevo G2S (Waters, Manchester, UK). Data were processed by using MassLynx V4.1 software.

\subsection*{F. Synthesis}
The detailed synthetic protocol used to prepare all molecules can be found in supplementary information.
\section*{Results and Discussion}

Several diverse fragments binding the viral M$^{pro}$ have been identified by high-throughput crystallographic screening of this protease. Among the screened fragments, x0195 (PDB ID: 5R81 \cite{douangamath-2020} - Figure 2A) shows one of the highest binding affinities \cite{kantsadi-2021} and therefore provides a reasonable starting point for fragment-based design of novel M$^{pro}$ inhibitors.

The crystal structure shows that x0195 is located within the M$^{pro}$ substrate binding pocket, at the interface of the two subpockets S2 and S4 as described by Cannalire, et al. \cite{cannalire-2020}. S4 is a solvent exposed subpocket that is partially composed by a flexible loop delimited by Gln189 and Gln192, while S2 is defined by the side chain residues of Phe140, Asn142, His163, Glu166, and His172, and the backbone atoms of Phe140 and Leu141. 

In the co-crystal structure corresponding to M$^{pro}$ in complex with x0195 (see Figure 2A), the aromatic portion of the molecule is located between the side chains of Gln189 and Met 165, while the unsaturated region of the tetrahydroquinoline scaffold establishes a hydrophobic interaction with the side chains of His41 and Met49. The N-methyl group attached to the tetrahydroquinoline core is solvent exposed, while the sulfonamide moiety is in contact with Pro168 and Glu166. In particular, the aromatic ring of the small molecule is bisecting the SO2 unit and the polar sulfonamide nitrogen (-NH2) is reaching the boundaries of the hydrophobic part of the binding pocket composed by the alkyl chain of Pro168. 

After comparing the available X-ray structural information with previously conducted studies on small molecule conformational preferences derived from crystal structure data \cite{brameld-2008}, we noticed that x0195 was modeled in a high energy conformation and that an unusual high-energy (i.e. repulsive) contact occurs between the sulfonamide oxygen and the carbonyl oxygen of Glu166 backbone. Additionally, the tetrahydroquinoline scaffold was not fully exploring S2 subpocket boundaries. As reported by Cannalire, et al. \cite{cannalire-2020} and Zhang, et al. \cite{zhang-2020}, the volume of the S2 subpocket in SARS-CoVs M$^{pro}$ is highly similar to that of the MERS-CoV homologue. However, the volume of S2 in SAR-CoVs M$^{pro}$ (252 Å$^{3}$) is significantly larger than in other CoVs’ homologues of the $\alpha$-genus, such as the HCoV-NL63 M$^{pro}$ (45 Å$^{3}$)\cite{zhang-2020, cannalire-2020}. Therefore, exploiting this knowledge might be key to design specific inhibitors of CoVs M$^{pro}$.

In order to refine the available X-ray structural model and to gather more structural insights (e.g. protein flexibility and binding pocket rearrangements \cite{D1SC00145K,acs.jpclett.1c01460}) to guide the design of better binders of the subpocket S2, we ran all-atom molecular dynamics simulations using the AMOEBA polarizable force field \cite{ren2003polarizable, ponder2010current, shi2013polarizable, AMOEBAnuc}  on M$^{pro}$ (PDB code: 7L11) in complex with x0195 (PDB code: 5R81).

\begin{figure*}[ht]
  \centering
    \includegraphics[width=0.9\textwidth]{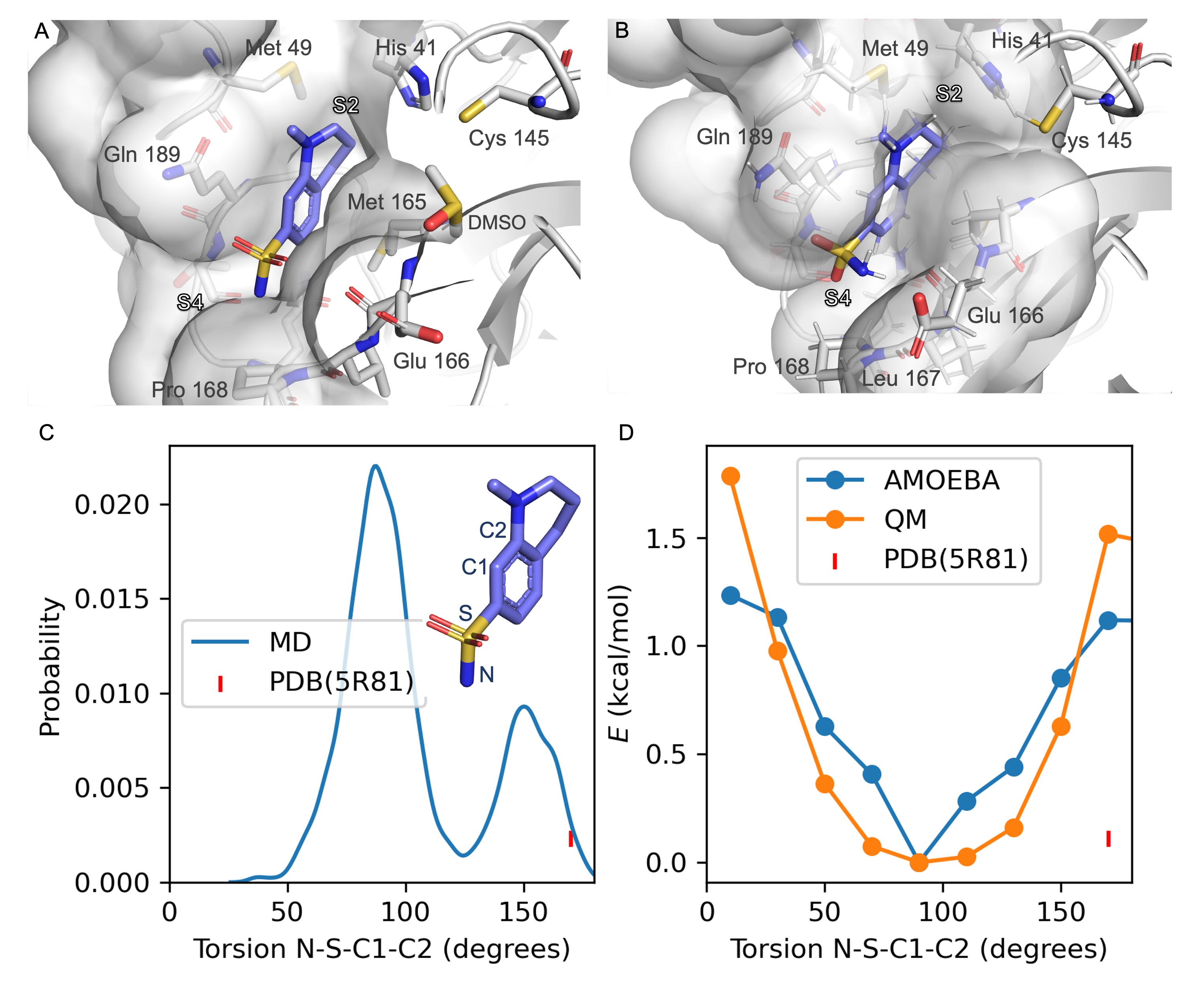}
  \label{Fig:Figure1}
  \caption{Refinement of the co-crystal structure of x0195 and M$^{pro}$ using MD simulations. A) An unusual conformation, of x0195 (in purple) located in the binding pocket formed by His 41, Met 49, Glu 166, Gln 189, and Pro 168 and their surroundings (PDB code: 5R81), and B) the relaxed structure of x0195 (in purple), obtained after the equilibration step, interacting with the amino acid residues of the substrate binding site. M$^{pro}$ is shown in light grey. C) Torsion angle distribution for the sulfonamide group during 20 ns of MD simulations (in blue) performed on M$^{pro}$ dimer in complex with x0195 ; the torsion angle of the sulfonamide group in the co-crystal structure is shown in pink. D) Torsion energy scan calculated by AMOEBA (in blue) and QM (in orange); the torsion angle of the sulfonamide group in the co-crystal structure is shown in pink. QM level=$\omega$B97x-D/6-31g* \cite{b97xd,631G,g16}} 
  \end{figure*}

Our simulations show that the unusual high-energy contacts between the sulfonamide oxygen and the carbonyl oxygen of Glu166 backbone were released. Also, regarding the electronic structure, we noticed that the p orbitals of the aromatic carbon C1 bisect (e.g. are parallel to) the SO2 angle, compared with a 90° value for the same angle as reported in the crystal structure (see Figure 2). Moreover, the NH2 of the sulfonamide group is engaging in favorable polar interactions with the Gln189 side chain and the solvent. 

Then, we performed absolute binding free energy calculations on the refined protein-ligand structure. Our results show that x0195 binds to the protein with a binding free energy of  -2.83 kcal/mol at 283 K, which is comparable to the experimental binding energy (-3.59±0.1 kcal/mol, see Table \ref{Tab:deltag}).

\begin{table}[!h]
\begin{tabular}{ccc}
\hline
 Compound    & Computed $\Delta G$ & Experimental $\Delta G$  \\
 \hline
QUB-00006 (R) & -2.73 $\pm$ 0.34 & \multirow{2}{*}{N.A.} \\

QUB-00006 (S)& -2.72 $\pm$ 0.22 & \\ 

QUB-00006-Int-01 (R) & -4.30 $\pm$ 0.35 & \multirow{2}{*}{-3.71 $\pm$ 0.2}   \\

QUB-00006-Int-01 (S)& -4.45$ \pm$ 0.29 &     \\

x0195 & -2.83$\pm$ 0.66 & -3.59 $\pm$ 0.1 \\
QUB-00006-Int-07 &-5.37$\pm$0.23 & covalent binder \\
\hline
\end{tabular}
\caption{Experimental and computed  
binding free energies (kcal/mol) for the non-covalent compounds. N.A.=not available (see text for details). }
\label{Tab:deltag}
\end{table}

\begin{figure*}[h]
  \centering
     \includegraphics[width=0.95\textwidth]{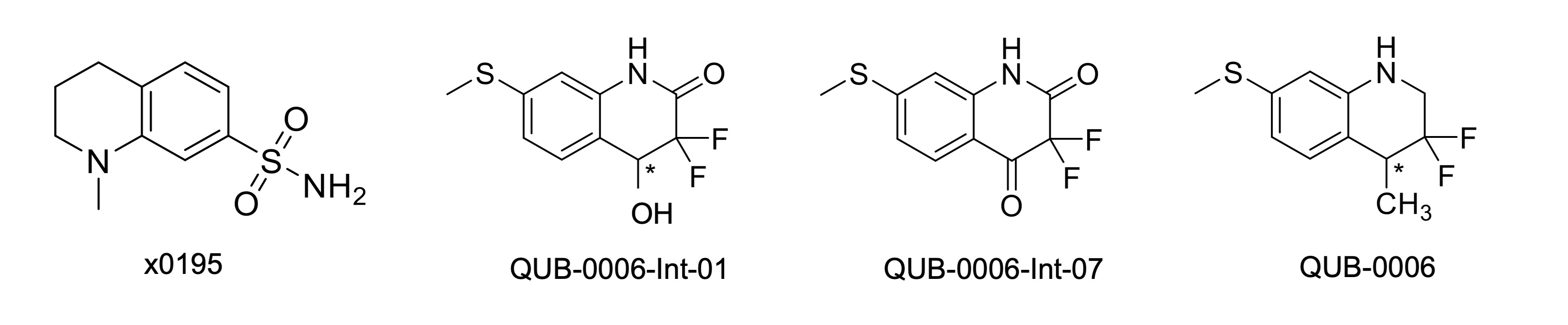}
  \label{Fig:Figure0}
  \caption{2D structures of: A) x0195; B) QUB-00006-Int-01; C) QUB-00006-Int-07, and, D) Qub-00006. The asterisk represents a chiral center.} 
\end{figure*}

We obtained the experimental binding free energy by converting the experimental Kd (1.7 mM ± 0.2) provided in literature \cite{kantsadi-2021} using the Gibbs free energy equation and the experimental temperature used in the binding assays (283 K). The agreement of the computed free energy prediction with the experimental results is reasonable. Further analysis of MD simulations suggests that the tetrahydroquinoline scaffold of x0195 is sub-optimally occupying the binding pocket. 

We put in place design strategies to modify the chemical moieties of x0195 and potentially increase its binding affinity. Here, we introduce the design of a new molecule, namely QUB-00006, where we added two fluorines and a methyl group on the tetrahydroquinoline core of x0195. Also, we substituted the sulfonamide group on the aromatic ring of the molecule by a methanethiol. Fluorination at position 3 of the tetrahydroquinoline core could increase ligand occupancy with no disruption of the water network surrounding the binding pocket \cite{D1SC00145K,acs.jpclett.1c01460}, while methylation at position 4 seemed an interesting modification to increase the potential interactions of the ligand with binding pocket residues.  
QUB-00006 was generated based on the structure and position of x0195 in the co-crystal (5R81), then placed in the receptor structure (M$^{pro}$ dimer with the PDB code: 71LL); next, the complex M$^{pro}$-QUB-00006 was equilibrated using MD simulations (see SI, section 1 for the detailed protocol), followed by free energy calculations.
To explore the potential of our computational platform in designing new binders with no or few experimental data such as ligand-M$^{pro}$ co-crystal structures, we leveraged all-atom molecular dynamics simulations on QUB-00006 complexed with M$^{pro}$. The aim of this approach is to gather insights on the binding conformation of the newly \textit{in silico} designed ligand, assess pocket fitness, and evaluate its binding affinity using ABFE calculations. 

The initial molecular conformation is mostly anchored at the binding pocket, with the $\alpha,\alpha$-difluoro-methyl group attached to the tetrahydroquinoline core fully occupying the buried part of the S2 subpocket, which is composed by the side chains of Met49 and His41, while the  sulfonamide moiety extends to S4 (Leu167 and Pro168). 
We note that methylation at position 4 of the tetrahydroquinoline core introduces a chiral center, however no significant differences in terms of pocket occupancy between the R and S enantiomers were observed.

The computed absolute binding free energies for QUB-00006(R) and QUB-00006(S) are -2.73±0.34 kcal/mol and -2.72±0.22 kcal/mol, respectively (Table \ref{Tab:deltag}). These results suggest that the designed fluorinated fragment is a binder at the M$^{pro}$ S2 subpocket and could represent a starting point for structure-based design of novel M$^{pro}$ inhibitors.

The identified binding mode is defined by several favorable intermolecular interactions occurring between the newly designed ligand and the M$^{pro}$ binding pocket: i) the sulfur group of QUB-00006(R) interacts with the oxygen of the carbonyl belonging to the backbone of Glu166 with a distance of 3.3 Å, ii) the $\alpha,\alpha$-difluoro moiety points towards the His41, and iii) the sulfur of Met49 establishes a favorable interaction with one of the two fluorines of the substrate (distance 3.3Å). 
In fact, the sulfur-oxygen contact observed in our simulations is in agreement with the findings of a study conducted by Iwoaka et al., \cite{iwaoka-2002} where they found that a total of 1200 and 626 fragments from the Cambridge Structural Database (CSD) and Protein Data Bank (PDB), respectively, have close intermolecular S-O contacts (with a distance of 3.52 Å or less). Another study analyzing the protein structures deposited in the Protein Data Bank reports 1133 interactions between His and halogen atoms found in 3833 PDB entries with one or more halogenated ligands co-crystallized with a protein \cite{kortagere-2008}. Moreover, the strong S-F interaction identified during the simulations is in good agreement with experimentally observed distances for fluorine-sulfur contacts in crystal structures (2.8-3.4 Å) \cite{bauer-2016}. It is worth noting that such interactions involving sulfur and halogen atoms are usually better captured with polarizable models than with their classical counterparts\cite{hage2013could,doi:10.1021/jp411671a,melcr2019accurate}.

\begin{figure*}[h]
  \centering
  \includegraphics[width=0.95\textwidth]{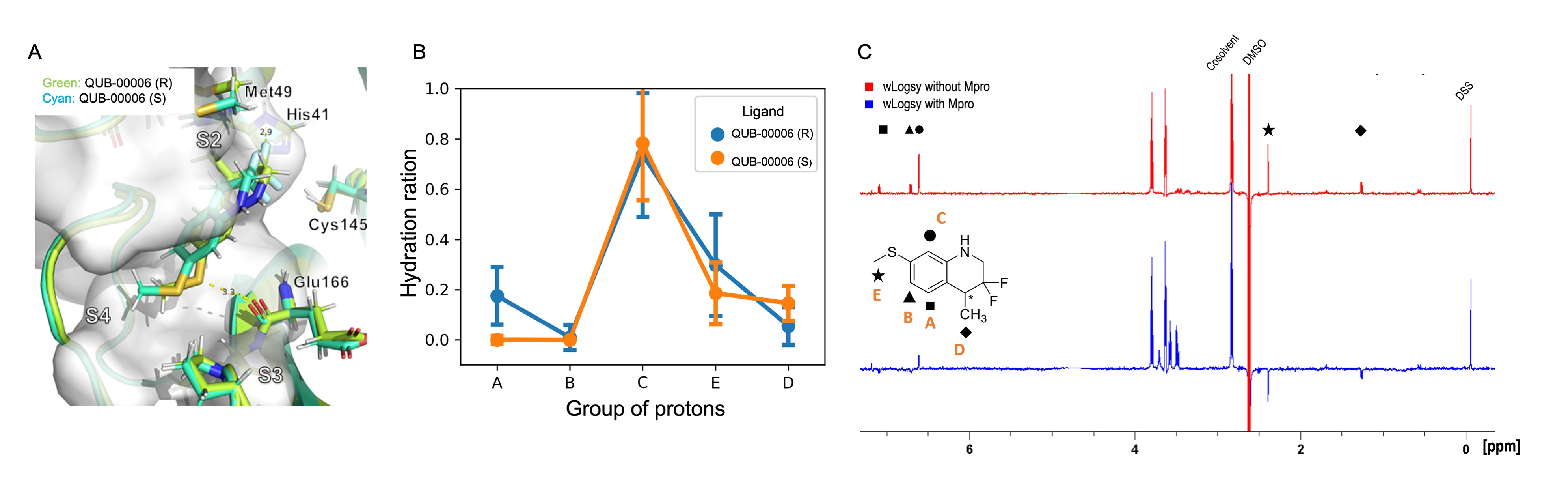}

  \caption{Computational and experimental characterization of QUB-00006 binding within M$^{pro}$ binding pocket. A) QUB-00006(R) (in light green) and QUB-00006(S) (in cyan) bind in a similar fashion at the interface of subpockets S2 and S4; the binding poses shown here were clustered and extracted from the trajectories of the binding free energy calculations performed on QUB-00006(R) and QUB-00006(S). B)  The analysis of our binding free energy trajectories shows that protons in group A, B, E, and D have a low hydration ratio (less than 0.5), while the proton of group C has a high hydration ratio of 0.8. Hydration ratios calculated for the different proton groups of QUB-00006(R) correlate with those calculated for QUB-00006(S). C) The waterLOGSY spectra of QUB-00006 in presence and in absence of the M$^{pro}$. The assignment scheme is reported along with the 2D structure of the ligand. The strong negative intensity of the signals of the hydrogens of groups A, D, and E suggests that they are orientated towards the protein, whereas the hydrogen atom in C is solvent exposed. These experimental findings confirm the hydration ratio calculated during our binding free energy simulations and described in panel B.}
  \label{Fig:fig_ds16}
  
\end{figure*}

QUB-00006 was then synthesized following the path in Figure 4 in order to validate in-vitro the simulation outcomes.

The ligand orientation in the MD simulations and the computed hydration ratio of the different atoms of QUB-00006 during ABFE simulations suggest that proton C is solvent exposed, while the protons of the methyl thioether group (group E) and the methyl group at position 4 of the tetrahydroquinoline core (group D) are buried (Figure \ref{Fig:fig_ds16}B).

\begin{figure*}[htbp]
  \centering
  \includegraphics[width=0.95\textwidth]{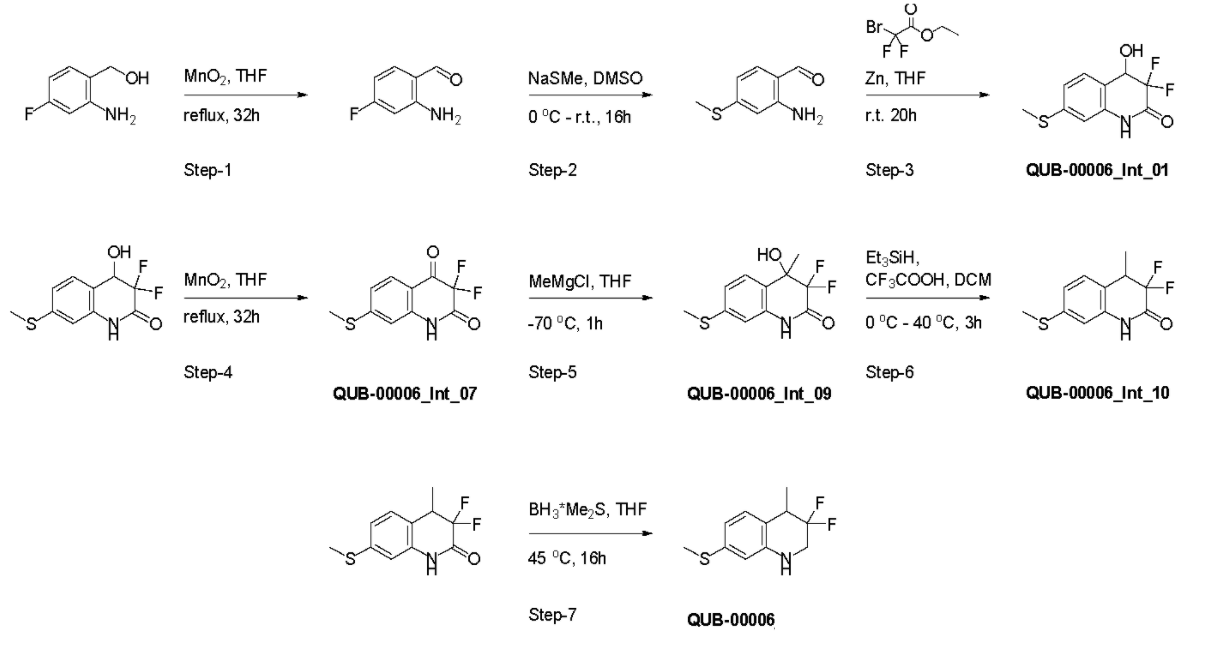}
  \label{Fig:fig_synthesis}
  \caption{Synthesis path of 3,3-difluoro-4-methyl-7-(methylsulfanyl)-1,2,3,4-tetrahydroquinoline named QUB-00006.}
  
\end{figure*}

Those findings strongly correlate with the NMR characterization of QUB-00006 obtained via WaterLogsy experiments. In fact, waterLogsy epitope mapping confirms that QUB-00006 binds to the protein binding pocket. We leveraged the experimental approach to better identify the region of the ligand in contact with the protein. In Figure \ref{Fig:fig_ds16}C, the proton signals arising from the two methyl groups (D ad E) in presence of M$^{pro}$ show a change in the sign suggesting that these protons are in close contact with the protein. Similarly, the aromatic protons A and B, undergo a sign inversion. On the contrary, the aromatic proton C is not significantly perturbed, which suggests this position is solvent exposed. The binding mode suggested by NMR is in agreement with the MD-derived hydration ratios confirming the predictive power of our MD-based approach to characterize the binding mode of novel ligands with experimental level of accuracy (Figure \ref{Fig:fig_ds16}B and C).

Although we were able to gather structural information about the binding mode of QUB-00006 using WaterLogsy assay, we could not measure its experimental binding affinity via STD NMR due to solubility challenges. 

Several synthetic steps were performed in order to obtain QUB-00006, as detailed in Figure 4. Through this synthetic scheme, we obtained different intermediates characterized by a better solubility profile (Table \ref{Tab:solubility}). Interestingly, the hydroxyquinolinone QUB-00006-Int-01 displayed the best solubility profile of all the synthetic intermediates, making it a strong candidate for in-vitro evaluation.

\begin{table}[!h]
\begin{tabular}{cccccc}
   & MW (Da) & logS      &  logP       & \makecell{Tanimoto\\ (MACCS)} \\
QUB-00006  & 229.07 & -3.99 & 3.56 & 0.391\\
QUB-00006-Int-07 & 243.02  &  -3.73 & 1.96 & 0.371 \\
QUB-00006-Int-01 & 245.03 &   -2.73 & 1.66 & 0.338 \\
x0195 & 226.08  & -1.94 & 0.56 & 1 \\
\end{tabular}
\caption{\label{Tab:solubility}. Prediction of compounds properties using our Machine Learning workflow. MW represents the molecular weight of the compounds in daltons, logS is the predicted solubility of the different compounds, logP represents the differential solubility, and the Tanimoto coefficient reflects the similarity of the selected compounds relatively to x0195.}
\end{table}

Before conducting NMR STD experiments to determine the dissociation constant (Kd) of the more polar QUB-00006-Int-01 compound, we decided to predict its binding conformation at the binding pocket and compute the respective absolute binding free energy. Modification of the molecular scaffolds, especially in fragment-like molecules, might affect the binding mode\cite{onescaffold} compared to a reference structure (e.g. x0195 as per PDB ID:5R81). 

We used a combination of docking, MD and ABFE calculations to explore the putative binding mode of QUB-00006-Int-01. Those calculations identified two dominant binding modes for QUB-00006-Int-01(R) and QUB-00006-Int-01(S) (Figure 5A) with computed binding free energies of -4.4 and -4.3 kcal/mol, respectively. Then, we estimated the binding affinity of QUB-00006-Int-01 towards M$^{pro}$ by a STD NMR titration and we found a dissociation constant in the low millimolar range, with an estimated Kd of 1.9 $\pm$ 0.6 mM (-3.71 ± 0.2 kcal/mol), which agrees reasonably well with our binding free energy calculations (Table \ref{Tab:deltag}). As shown on Figure 5A, both enantiomers bind to the S2 and S4 subpockets with the thioether group being fully buried in subpocket S2, which correlates with WaterLogsy experiments (Figure 5C). Additionally, QUB-00006-Int-01(R) and QUB-00006-Int-01(S) fill up a binding pocket space that is different from the one occupied by QUB-00006. On the other hand, starting with a QUB-00006-like binding mode, we ran an additional absolute binding free energy calculation on M$^{pro}$:QUB-00006-Int-01R complex and obtained a binding free energy of -0.9 kcal/mol. These results suggest that QUB-00006-Int-01 and QUB-00006 might have different dominant binding conformations (see Figure 3A and Figure 5A).

\begin{figure*}[htbp]
  \centering
  \includegraphics[width=0.95\textwidth]{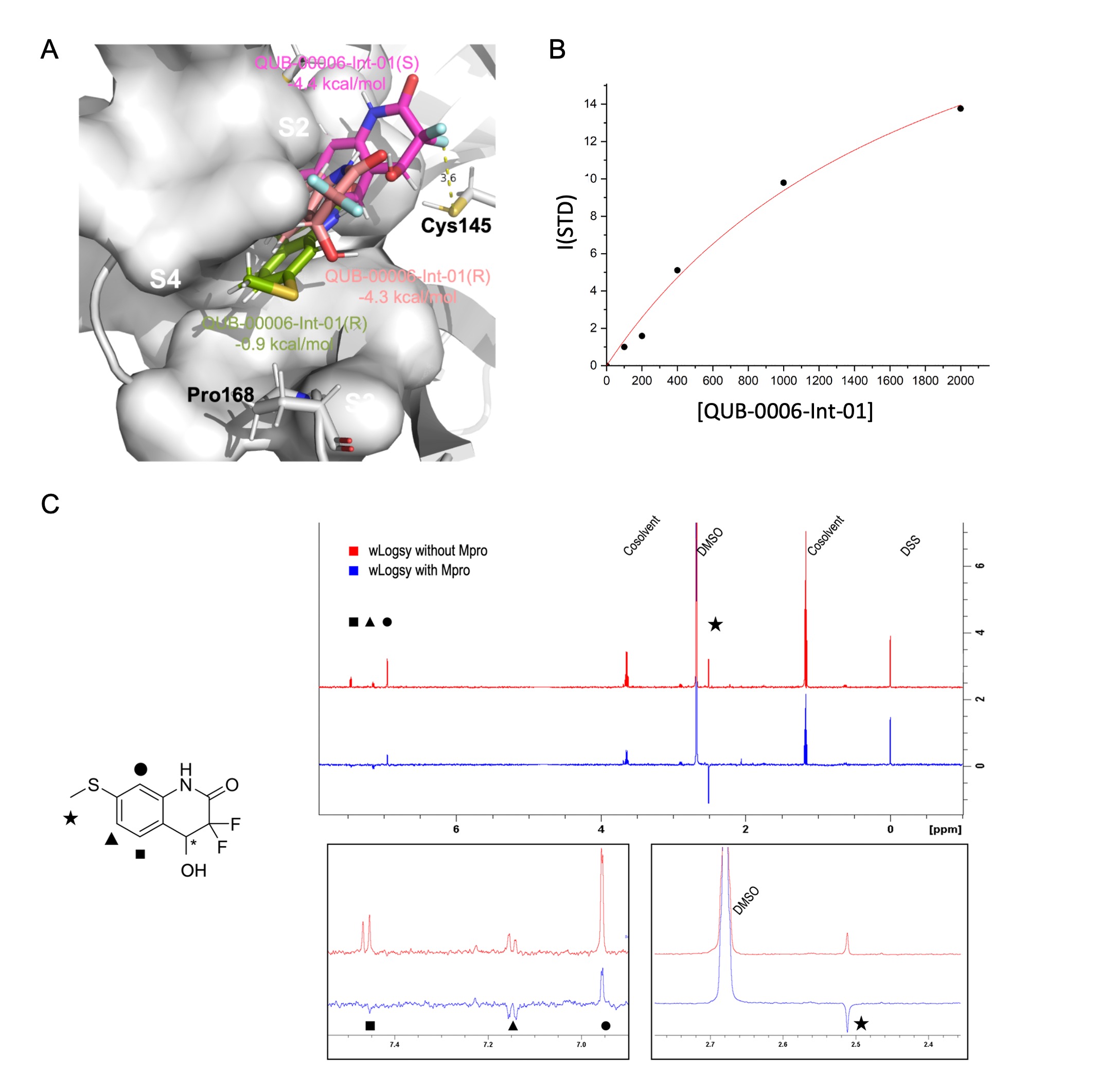}
  \label{Fig:fig_ds7_abfe}
  \caption{Computational and experimental characterization of QUB-00006-Int-01 in M$^{pro}$ binding pocket. A) The dominant binding modes of QUB-00006-Int-01(R) (in pink) and QUB-00006-Int-01(S) (in magenta), identified during ABFE simulations, have computed binding free energies of -4.4 and -4.3 kcal/mol, respectively; also, they bind to the S2 and S4 subpockets in a similar fashion with the thioether group being fully buried in S2. On the other hand, starting with a QUB-00006 like binding mode, we ran an additional absolute binding free energy calculation on M$^{pro}$ in complex with QUB-00006-Int-01(R) and obtained a second binding mode for QUB-00006-Int-01(R) (in green) with a binding free energy of -0.9 kcal/mol. B) STD titration profile of QUB-00006-Int-01. The ligand concentration ranges from 100 $\mu${M} to 2 mM against 10 $\mu${M} of M$^{pro}$. C) The WaterLOGSY spectrum of QUB-00006-Int-01 with M$^{pro}$ (in blue) and without Mpro (in red). The assignment of the signals is reported on the 2D structure of the fragment. The methyl and the aromatic signals of the two protons adjacent to the hydroxyl group undergo a significant change, which suggests that these groups are in close contact with the protein's cavity. On the contrary, the aromatic proton adjacent to the lactamic nitrogen undergoes a reduction of its intensity, suggesting that this proton is partially exposed to the solvent. These STD results confirm our computational characterization of the binding mode of QUB-00006-Int-01 (Panel A).}
   
\end{figure*}

Since a fragment-like molecule could have multiple binding modes and the ligand conformation is unlikely to be fully sampled during 20 ns of binding free energy simulations, we used unsupervised adaptive sampling (AS) to further explore the conformational space of QUB-00006-Int-01. AS can be use here as an interpretative tool able to gather structural insights on the various potential M$^{pro}$-ligand interactions (see SI for details). The AS trajectories were clustered using average-linkage hierarchical clustering algorithms and the top ten largest clusters were chosen for analysis. These clusters have comparable populations (the smallest clusters have 3-4 times smaller populations or 0.3 kcal/mol higher free energy than the largest clusters, see Table \ref{Tab:Table-S1}), indicating the coexistence of multiple binding modes.

\begin{table}[]
\begin{tabular}{rrrrrr}
\multicolumn{3}{c}{QUB-00006-Int-01(R)} & \multicolumn{3}{c}{QUB-00006-Int-01(S)}   \\
cluster     & Fraction & $\Delta\Delta G$      & cluster     & Fraction  & $\Delta\Delta G$      \\
1 & 0.101 & 0 & 1 & 0.103 & 0 \\
2 & 0.083 & 0.05 & 2 & 0.093 & 0.03 \\
3 & 0.067 & 0.11 & 3 & 0.088 & 0.04 \\
4 & 0.053 & 0.17 & 4 & 0.065 & 0.12 \\
5 & 0.042 & 0.23 & 5 & 0.059 & 0.14 \\
6 & 0.035 & 0.27 & 6 & 0.054 & 0.17 \\
7 & 0.034 & 0.28 & 7 & 0.049 & 0.19 \\
8 & 0.033 & 0.29 & 8 & 0.039 & 0.25 \\
9 & 0.032 & 0.30 & 9 & 0.033 & 0.29 \\
10 & 0.032 & 0.30 & 10 & 0.031 & 0.31
\end{tabular}
\caption{\label{Tab:Table-S1}Population of the clusters generated by adaptive sampling performed on M$^{pro}$ in complex with QUB-00006-Int-01 (R) and (S). $\Delta\Delta G$ (kcal/mol) is the relative free energy at 298 K. The relative binding free energies reported for QUB-00006-Int-01 (R) and (S) are calculated using the respective cluster 1 as a reference ligand.}
\end{table}
More precisely, starting from these clusters, absolute binding free energies would yield results within 0.3 kcal/mol of what was previously obtained. The simulations of QUB-00006-Int-01(R) and QUB-00006-Int-01(S) converged to similar ensembles containing several possible binding modes. Clusters 3, 5, and 6 of QUB-00006-Int-01(R) and cluster 4 of QUB-00006-Int-01(S) SI-Figure 2) correspond to the respective dominant binding modes predicted by ABFE simulations (Figure 5A). For both enantiomers, the most conserved interactions are the hydrophobic contacts between C9 (methyl thioether) and Gln189, and between C5 (proton B) with His41, Arg188, and Gln189.

Overall, our computational findings on QUB-00006-Int-01 confirm that the structural approach we introduce in this work using a sequence of MD-based techniques (classic MD simulations, adaptive sampling, and absolute binding free energy calculations) is able to capture potential binding orientations of fragment-like compounds in the binding pocket of a protein, and to accurately predict their binding free energies.

\begin{figure}[h]
  \centering
  \includegraphics[width=0.9\linewidth]{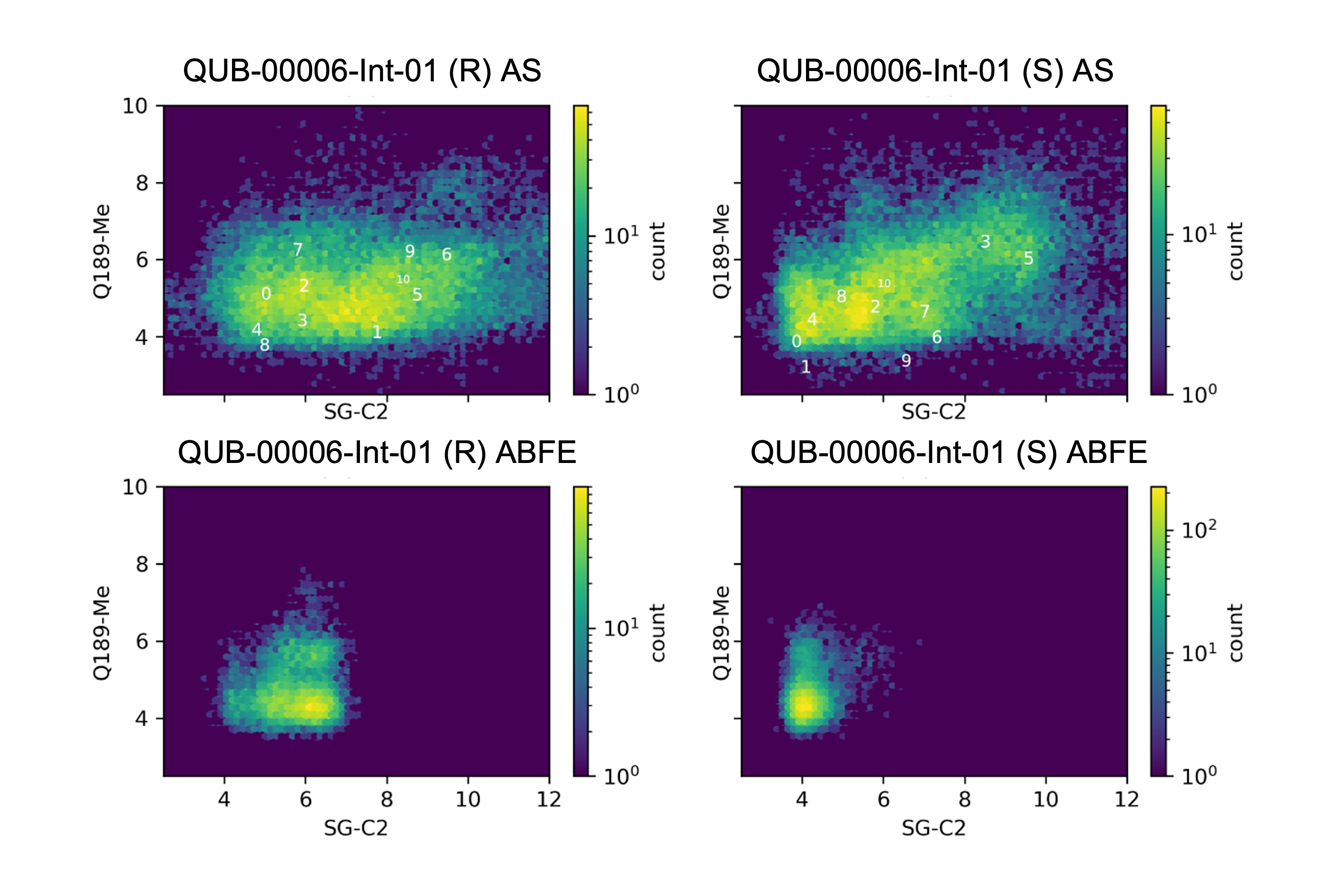}

  \caption{Conformations of QUB-00006-Int-01 sampled during 20 ns of ABFE calculations and 450 ns of adaptive sampling simulations. The conformation was plotted as a function of two distances: i) the distance between C2 (carbon of QUB-00006-Int-01 connected to the hydroxyl group) and the sulfur of Cys145, and ii) the distance between the methyl thioether group in QUB-00006-Int-01 and the beta carbon of Gln189. “0” indicates the starting structure, “1” indicates the largest cluster, and “i” indicates the ith largest cluster. The frames were taken at 10 ps time interval.}
    \label{Fig:fig_QUB-00006-Int-01_hm}
\end{figure}

Then, we analyzed the clustered QUB-00006-Int-01 binding conformations from the adaptive  sampling simulations plotted as a function of the distance between the methyl thioether group in QUB-00006-Int-01 and the beta carbon of Gln189, and the distance between C2 (carbon connected to hydroxyl) and the sulfur atom (SG) of the catalytic side chain of Cys145 residue (Fig. \ref{Fig:fig_QUB-00006-Int-01_hm}). We noticed that the distance of C2-SG in the most populated cluster generated by the AS simulations is around 4 Å. 
To reinforce our analysis, we leveraged another unsupervised reduction of dimension technique: TICA (the time-lagged independent component analysis)\cite{perez2013identification}, which aims at finding the slow collective variables of the data, and applied it to QUB-00006-Int-01(R). We then used the k-mean clustering method on the data projected on this space and built a Hidden Markov State Model (HMSM)\cite{scherer2015pyemma}. Three clusters emerged, whose characteristics also show the coexistence of several binding modes of QUB-00006-Int-01(R), one of which corresponding to a distance between C2 and SG below 4 Å. Detailed results can be found in SI. 

\begin{figure}[h!]
  \centering
  \includegraphics[width=0.9\linewidth]{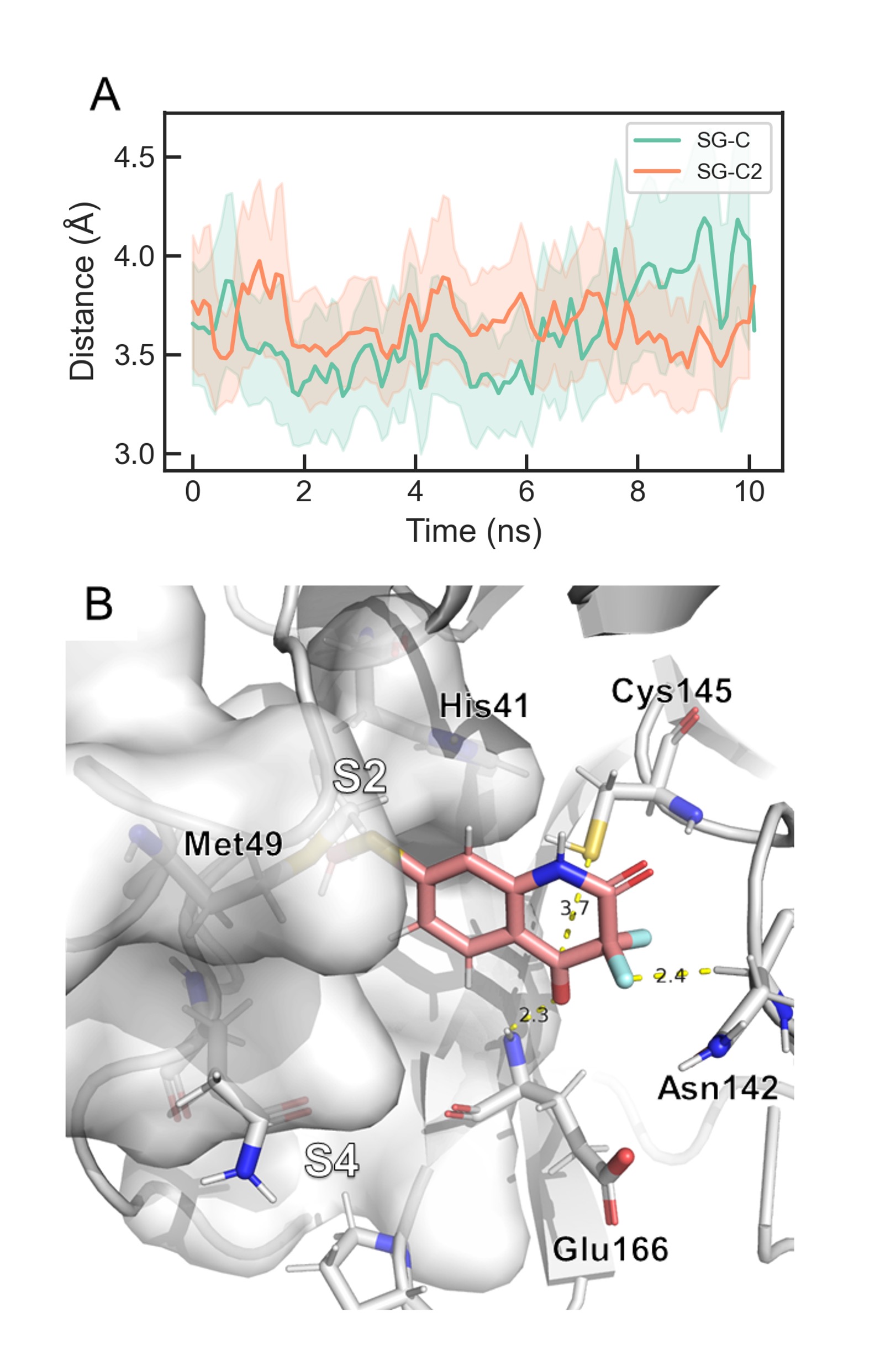}
  
  \caption{The dominant binding mode of QUB-00006-Int-07 during ABFE simulations. A) Time evolution of key distances in the simulation. "SG" stands for the sulfur atom in Cys145, "C" is the amide carbon in QUB-00006-Int-07, and "C2" is the carbonyl carbon in QUB-00006-Int-07. The average distances for SG-C and SG-C2 are 3.61 Å and 3.65 Å, respectively. B) The dominant binding mode of QUB-00006-Int-07 within M$^{pro}$ binding pocket. QUB-00006-Int-07 is shown in pink and the protein is shown in silver sticks and surfaces. The binding mode is very stable during the simulation, where the hydroxyl group is close to Cys145 and forms a hydrogen-bond with Glu166 backbone, and the difluoro group interacts with the carbonyl group of Asn142. These binding modes are also comparable to the dominant binding mode of QUB-00006-Int-01(S) identified during ABFE calculations.}
  \label{Fig:Figure5b}
  
\end{figure}

Targeting Cys145 with covalent warheads has been used by several researchers to discover novel potent inhibitors of M$^{pro}$ \cite{cannalire-2020,Arafet2021,marti2021impact}. As a matter of fact, a simple chemical modification to QUB-00006-Int-01 would lead to QUB-00006-Int-07 bearing an $\alpha,\alpha$-difluoro-keto moiety, which is prone to a nucleophilic attack by the vicinal R-SH of Cys145. In order to enable the latter, QUB-00006-Int-07 would need to access the M$^{pro}$ substrate pocket and adopt a stable binding conformation prior to the covalent binding to occur. Thus, we conducted absolute binding free energy simulations on the M$^{pro}$:QUB-00006-Int-07 complex, which confirmed a favorable binding energy of QUB-00006-Int-07 to the M$^{pro}$ substrate pocket (-5.37$\pm$0.23 kcal/mol). As reported in Figure 7, compound QUB-00006-Int-07 is bound to the S2 and S4 subpockets with the thioether group being fully buried in subpocket S2 and the $\alpha,\alpha$-difluoro-keto moiety facing Cys145. More precisely, the average distance between SG and the C is 3.65 Angstroms ($\pm$ 0.33) and the average distance between the SG and C2 is 3.61 Angstroms ($\pm$ 0.43) as can be seen in Figure \ref{Fig:Figure5b}. 

Our computational findings motivated us to test the compound with a FRET-based proteolytic assay. This assay should detect potent functional binders to the viral Mpro. Being a fluorogenic assay, compounds with fluorescence quenching properties can suppress the fluorescence signal generated by the protease activity. To eliminate false positive results, we conducted a preliminary counter screen and verified that the tested compound possesses negligible fluorescence quenching effects. Subsequently, to assess the potential inhibitory activity of the compound against SARS-CoV-2 M$^{pro}$, increasing concentrations of QUB-00006-Int-07 (0.25-150 uM) were incubated with 20 nM M$^{pro}$ before the addition of 5 $\mu$M FRET substrate. As shown on Figure 8, QUB-00006-Int-07 inhibited M$^{pro}$ with 50\% inhibitory concentration (IC50 value of 830 ± 50 nM), thus resulting in a fairly potent inhibitor of the M$^{pro}$ enzymatic activity. The binding of QUB-00006-Int-07 to M$^{pro}$ was confirmed by electrospray ionization (ESI) mass spectrometry.

\begin{figure}[h]
  \centering
  \includegraphics[width=0.9\linewidth]{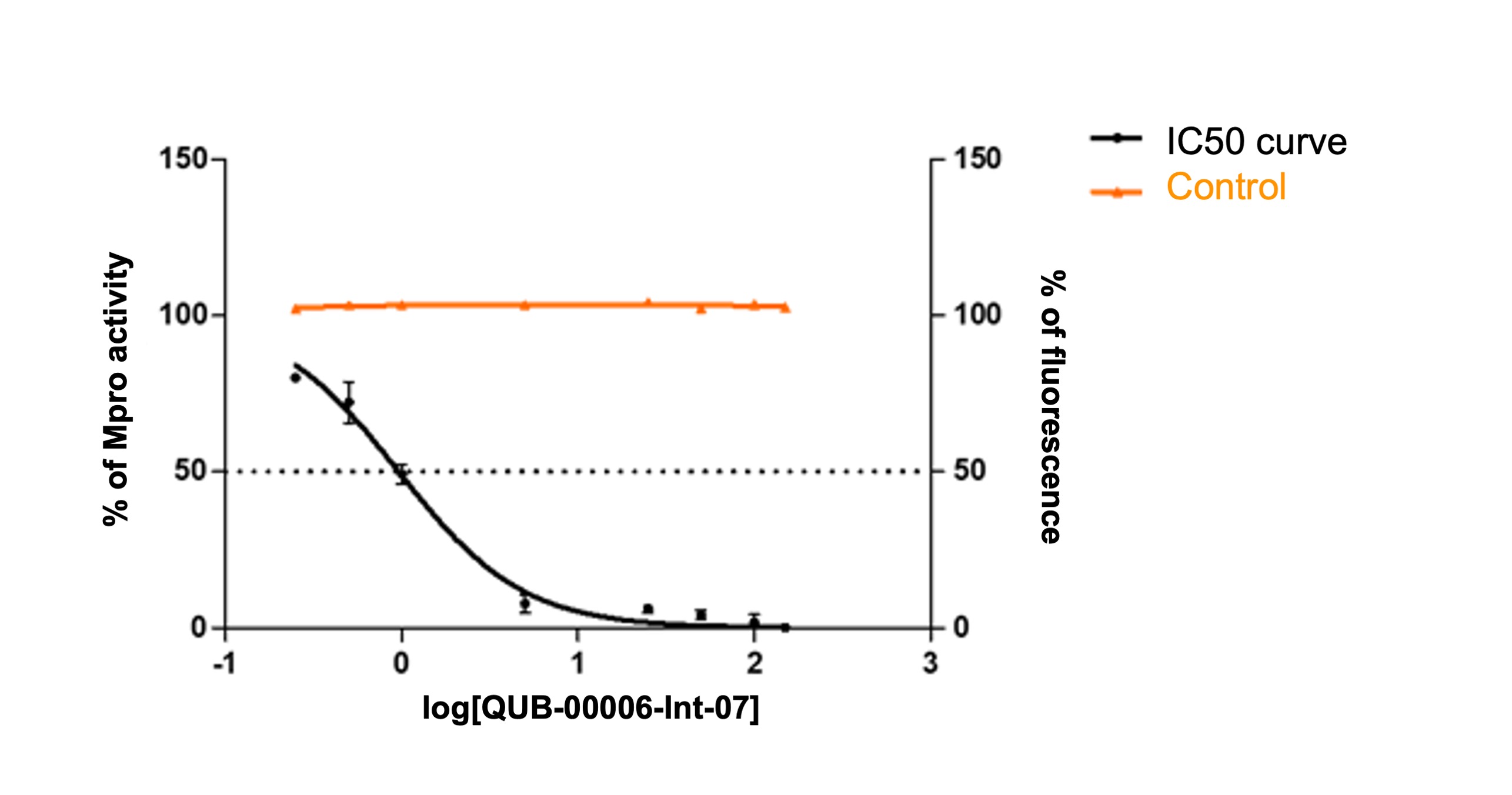}
  \label{Fig:Figure9}
  \caption{Dose-response curves obtained by plotting the percentage of SARS-CoV-2 M$^{pro}$ residual activity as a function of increasing concentrations of QUB-00006-Int-07 (0-150 µM). [ M$^{pro}$]= 20 nM, [PS1]= 5 µM, \%DMSO=3.75\%. Experiments were performed in triplicate. A counter screening control experiment was performed by testing increasing concentration of QUB-00006-Int-07 in the presence of 0.5 µM free fluorescein.
}
  
\end{figure}

A preliminary determination of the initial protein showed an experimental mass of 33796.40 Da, which matches very closely the expected value of 33796.64 Da calculated from the sequence (Figure 9A). The sample obtained after incubation of QUB-00006-Int-07 with M$^{pro}$ (compound:protein=10:1 ratio) was analyzed by ESI-MS in denaturing conditions, and a representative spectrum is provided in Figure 9B. In addition to the signals corresponding to multiple charge states of the initial protein (red dots), we identified the distribution of signals corresponding to the M$^{pro}$ modified by the presence of the compound (green asterisks) which is therefore covalently linked to the protein given the non-native conditions of the experiment. The nature of the adduct and the molecular mechanism of binding is under investigation and will be subject of further studies. 

Finally, in this work, the introduction of multiple modifications (e.g. gem-difluoro, thioether, hydroxyl and methyl groups) to the tetrahydroquinoline scaffold of x0195, and the design and synthesis of novel molecular scaffolds, (see Table \ref{Tab:solubility}) enabled exploration of binding pocket boundaries and provided additional information related to druggability of the S2 subpocket. 
Other molecules were produced over the course of this research but, due to their weaker activity, their detailed analysis is not provided here. Their list can be found in SI. These compounds were either designed computationally without leading to improved affinities or were synthesis intermediates. All resulting molecules were submitted to biological testing, but none of them were found to be as potent as QUB-00006-Int-07 nor presented a strong druggable profile, compared to the previously discussed compounds. 

\begin{figure}[h]
  \centering
   \includegraphics[width=0.9\linewidth]{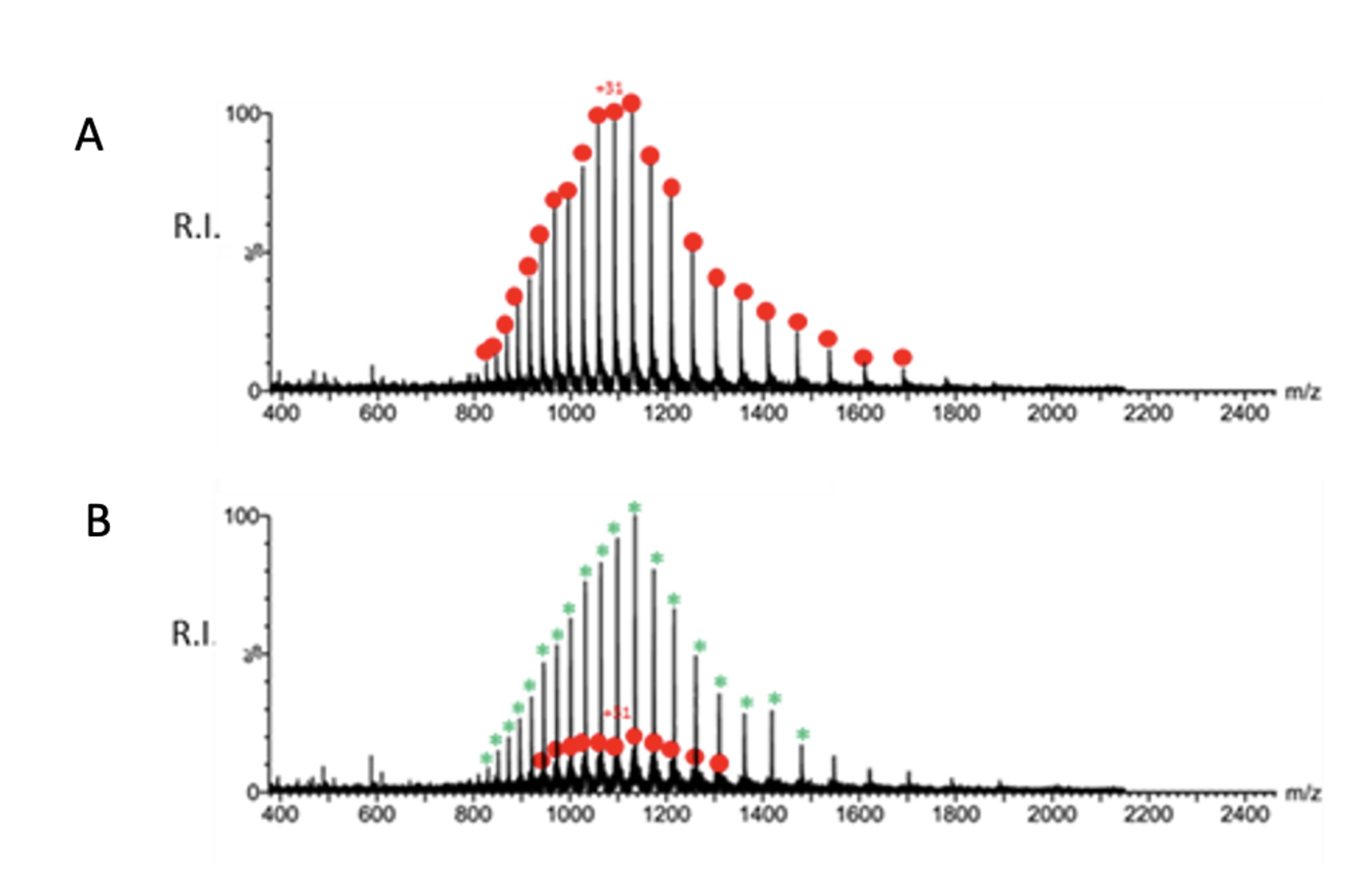}
  \label{Fig:Figure8}
  \caption{A) Representative ESI-MS spectrum of a solution containing 10 µM SARS-CoV-2 M$^{pro}$ in water/acetonitrile (50:50) added with 0.1\% formic acid. Spectrum was acquired in positive ion mode. B) Representative ESI-MS spectrum of a mixture containing 10 µM SARS-CoV-2 M$^{pro}$ after incubation with QUB-00006-Int-07 (compound:protein=10:1 ratio) in water/acetonitrile (50:50) added with 0.1\% formic acid. Spectrum was acquired in positive ion mode. The red dots correspond to the unmodified protein, the green asterisks correspond to the modified protein.}
  
\end{figure}

\section*{Conclusion and perspectives}
We presented a computationally-driven discovery of a new set of non-covalent and covalent inhibitors of M$^{pro}$ that have been further characterized experimentally. The best compound, QUB-00006-Int-07, has been found to be a covalent binder that resulted in a potent inhibition of the M$^{pro}$ enzymatic activity (IC50= 830 ± 50 nM). The results on the innovative scaffold design described here were obtained within three months via a fast track project that took place in the summer of 2021. It involved a small consortium of theoreticians, organic chemists and drug designers, and demonstrated the effectiveness of a computation-guided synthetic strategy. Indeed, GPU-accelerated high-performance computing platforms can now provide access to high-resolution molecular dynamics simulations, which are able to predict detailed protein conformational maps and to provide accurate absolute binding free energy results. Such computations can be further rationalized by means of adaptive sampling simulations, an approach which is able to decipher multiple binding modes. 
Coupled to NMR, in-vitro experiments and machine learning, such high-resolution predictions yield structural insights regarding the design of new active compounds, while offering an atomic level understanding of binding affinities. 

Beyond this preliminary proof of concept study, next research steps will be devoted to the QM/MM modeling \cite{loco1,loco2} of the warhead reaction mechanism \cite{D0SC02823A,Arafet2021,marti2021impact} leading to the covalent binding of QUB-00006-Int-07, and to optimization of active compounds with the goal of reaching low nanomolar activity.  

\section*{Author contributions statement}

L.E.K, Z. J., D.L, T.J.I., D.E.A, B.S, L.L performed simulations; \\
L.L., T.J.I., D.S, J-P.P contributed new code;\\
M.S, M.B, A.S., M. V. performed NMR and Mass spectroscopy experiments;\\
I.K., R.I., A.K., Y.M., A.P., V.C-R. performed synthesis; \\
L.L., B.S,P.R.,J.W.P., D.S, J-P.P contributed new methodology;\\
L.L., B.S, F.H.,P.R.,J.W.P.,J-P.P contributed analytic tools;\\
All authors analyzed data;\\
L.E.K, Z. J., L.L.,M.S.,A.S., D.S and J-P.P. wrote the paper;\\
D.S. and J-P.P designed the research. 
\section*{Conflicts of interest}
P.R., M.M., L.L., J.W.P. and J-P.P. are co-founders and shareholders of Qubit Pharmaceuticals.

\section*{Notes}
Qubit Pharmaceuticals and Sorbonne Université have submitted a preliminary patent application on the compounds reported in this study.

\section*{Acknowledgements}
This work has received funding from the European Research Council (ERC) under the European Union’s Horizon 2020 research and innovation program (grant agreement No 810367), project EMC2 (JPP). Computations have been made possible thanks to special COVID-19 grants from : i) Amazon Web Services (AWS) allowing to access the AWS Cloud supercomputing platform (JPP; Qubit Pharmaceuticals); ii) GENCI on the Jean Zay supercomputer (IDRIS, Orsay, France) through projects AP010712339 and AD011012316 (Qubit Pharmaceuticals); GENCI allocation no A0070707671 (JPP). The work  on experimental validation was supported by funding from the CARIPARO Foundation (“Progetti di ricerca sul Covid-19" N. 55812 to BG) and  from the Department of Chemical Sciences (project P-DiSC 01BIRD2018-UNIPD to MB). DS thanks Denis Klapishevskiy for discussions, Prof. Carsten Bolm and his team for providing the synthetic scheme for scaffolds related to x0195 fragment (https://bolm.oc.rwth-aachen.de/content/outreach). 
\scriptsize{
\bibliography{rsc} 
\bibliographystyle{rsc} } 

\end{document}


\pagestyle{fancy}
\thispagestyle{plain}
\fancypagestyle{plain}{
\renewcommand{\headrulewidth}{0pt}
}

\makeFNbottom
\makeatletter
\renewcommand\LARGE{\@setfontsize\LARGE{15pt}{17}}
\renewcommand\Large{\@setfontsize\Large{12pt}{14}}
\renewcommand\large{\@setfontsize\large{10pt}{12}}
\renewcommand\footnotesize{\@setfontsize\footnotesize{7pt}{10}}
\renewcommand\scriptsize{\@setfontsize\scriptsize{7pt}{7}}
\makeatother

\renewcommand{\thefootnote}{\fnsymbol{footnote}}
\renewcommand\footnoterule{\vspace*{1pt}%
\color{cream}\hrule width 3.5in height 0.4pt \color{black} \vspace*{5pt}} 
\setcounter{secnumdepth}{5}

\makeatletter 
\renewcommand\@biblabel[1]{#1}            
\renewcommand\@makefntext[1]%
{\noindent\makebox[0pt][r]{\@thefnmark\,}#1}
\makeatother 
\renewcommand{\figurename}{\small{Fig.}~}
\sectionfont{\sffamily\Large}
\subsectionfont{\normalsize}
\subsubsectionfont{\bf}
\setstretch{1.125} 
\setlength{\skip\footins}{0.8cm}
\setlength{\footnotesep}{0.25cm}
\setlength{\jot}{10pt}
\titlespacing*{\section}{0pt}{4pt}{4pt}
\titlespacing*{\subsection}{0pt}{15pt}{1pt}

\fancyfoot{}
\fancyfoot[LO,RE]{\vspace{-7.1pt}\includegraphics[height=9pt]{head_foot/LF}}
\fancyfoot[CO]{\vspace{-7.1pt}\hspace{13.2cm}\includegraphics{head_foot/RF}}
\fancyfoot[CE]{\vspace{-7.2pt}\hspace{-14.2cm}\includegraphics{head_foot/RF}}
\fancyfoot[RO]{\footnotesize{\sffamily{1--\pageref{LastPage} ~\textbar  \hspace{2pt}\thepage}}}
\fancyfoot[LE]{\footnotesize{\sffamily{\thepage~\textbar\hspace{3.45cm} 1--\pageref{LastPage}}}}
\fancyhead{}
\renewcommand{\headrulewidth}{0pt} 
\renewcommand{\footrulewidth}{0pt}
\setlength{\arrayrulewidth}{1pt}
\setlength{\columnsep}{6.5mm}
\setlength\bibsep{1pt}

\makeatletter 
\newlength{\figrulesep} 
\setlength{\figrulesep}{0.5\textfloatsep} 

\newcommand{\topfigrule}{\vspace*{-1pt}%
\noindent{\color{cream}\rule[-\figrulesep]{\columnwidth}{1.5pt}} }

\newcommand{\botfigrule}{\vspace*{-2pt}%
\noindent{\color{cream}\rule[\figrulesep]{\columnwidth}{1.5pt}} }

\newcommand{\dblfigrule}{\vspace*{-1pt}%
\noindent{\color{cream}\rule[-\figrulesep]{\textwidth}{1.5pt}} }

\makeatother

\twocolumn[
  \begin{@twocolumnfalse}

\vspace{1em}
\sffamily
\begin{tabular}{m{4.5cm} p{13.5cm} }

\includegraphics{head_foot} & \noindent\LARGE{\textbf{Computationally driven discovery of targeting SARS-CoV-2 Mpro inhibitors: from design to experimental validation$^\dag$}} \\
 & \vspace{0.3cm} \\

 & \noindent\large {Léa El Khoury,\textit{$^{a}$}\textit{$^{\ddag}$}
 Zhifeng Jing,\textit{$^{a}$}\textit{$^{\ddag}$}
 Alberto Cuzzolin,\textit{$^{b}$} 
 Alessandro Deplano,\textit{$^{c}$}
 Daniele Loco,\textit{$^{a}$}
 Boris Sattarov,\textit{$^{a}$}
 Florent Hédin,\textit{$^{a}$}
 Sebastian Wendeborn,\textit{$^{d}$}
 Chris Ho,\textit{$^{a}$}
 Dina El Ahdab,\textit{$^{o}$}
 Theo Jaffrelot Inizan,\textit{$^{o}$}
 Mattia Sturlese,\textit{$^{g}$}
 Alice Sosic,\textit{$^{e}$}
Martina Volpiana,\textit{$^{e}$}
Angela Lugato,\textit{$^{e}$}
Marco Barone,\textit{$^{e}$}
Barbara Gatto,\textit{$^{e}$}
Maria Ludovica Macchia,\textit{$^{e}$}
Massimo Bellanda ,\textit{$^{f}$}
Roberto Battistutta,\textit{$^{f}$}
Cristiano Salata,\textit{$^{h}$}
Ivan Kondratov,\textit{$^{i}$}
Rustam Iminov,\textit{$^{i}$}
Andrii Khairulin,\textit{$^{i}$}
Yaroslav Mykhalonok,\textit{$^{i}$}
Anton Pochepko,\textit{$^{i}$}
Volodymyr Chashka-Ratushnyi,\textit{$^{i}$}
Iaroslava Kos,\textit{$^{i}$}
Stefano Moro,\textit{$^{g}$}
Matthieu Montes,\textit{$^{l}$}
Pengyu Ren,\textit{$^{m}$}
Jay W. Ponder,\textit{$^{n}$}
Louis Lagardère,\textit{$^{o}$}
Jean-Philip Piquemal$^{\ast}$\textit{$^{o,p}$} and Davide  Sabbadin$^{\ast}$\textit{$^{a,d}$}} \\

\includegraphics{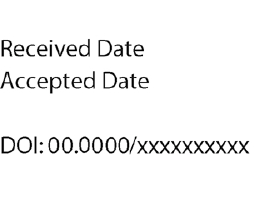} & \\

\end{tabular}

 \end{@twocolumnfalse} \vspace{0.6cm}

  ]

\renewcommand*\rmdefault{bch}\normalfont\upshape
\rmfamily
\section*{}
\vspace{-1cm}


\footnotetext{\textit{$^{a}$~QUBIT Pharmaceutical, Incubateur Paris Biotech Santé, 24 rue du Faubourg Saint Jacques
75014 Paris, France;  Contact email: davide@qubit-pharmaceuticals.com (DS)}}

\footnotetext{\textit{$^{b}$Current affiliation: Chiesi farmaceutici S.p.A, Nuovo centro ricerche, Largo belloli 11a, 43122, Parma (Italy)}}

\footnotetext{\textit{$^{c}$Current affiliation: Pharmacelera, Torre R, 4a planta, Despatx A05, Parc Científic de Barcelona, Baldiri Reixac 8, 08028 Barcelona, (Spain)}}

\footnotetext{\textit{$^{d}$University of Applied Sciences and Arts Northwestern Switzerland - School of LifeSciences, Hofackerstrasse 30, CH-4132 Muttenz, Switzerland }}

\footnotetext{\textit{$^{e}$Department of Pharmaceutical and Pharmacological Sciences, University of Padova, via Marzolo 5, 35131, Padova (Italy)}}

\footnotetext{\textit{$^{f}$Department of Chemistry, University of Padova, via Marzolo 1, 35131, Padova (Italy)}}

\footnotetext{\textit{$^{g}$Molecular Modeling Section, Department of Pharmaceutical and Pharmacological Sciences, University of Padua, via F. Marzolo 5, 35131, Padova, (Italy)}}

\footnotetext{\textit{$^{h}$Department of Molecular Medicine, University of Padua, via Gabelli 63, 35121, Padova, (Italy)}}

\footnotetext{\textit{$^{i}$Enamine LTD, 78 Chervonotkats‘ka str., Kyiv 02094 (Ukraine)}}

\footnotetext{\textit{$^{l}$Laboratoire GBCM, EA7528, Conservatoire National des Arts et Métiers, Hesam Université, 2 Rue Conte, 75003 Paris, France}}

\footnotetext{\textit{$^{m}$University of Texas at Austin, Department of Biomedical Engineering, TX 78705, USA}}

\footnotetext{\textit{$^{n}$Department of Chemistry, Washington University in Saint Louis, MO 63130, USA}}

\footnotetext{\textit{$^{o}$Sorbonne Université, Laboratoire de Chimie Théorique, UMR 7616 CNRS, 75005, Paris,France;Contact email: jean-philip.piquemal@sorbonne-universite.fr (JPP)}}
\footnotetext{\textit{$^{p}$Institut Universitaire de France, 75005, Paris,France.}}

\footnotetext{\textit{$^{\ddag}$These authors contributed equally to this work}}

\


{\LARGE{Supplementary Information\\}}
\vspace{\baselineskip} 
{
  \centering

\begin{figure*}[!htp]

\centering 
\includegraphics[width=1.0\linewidth]{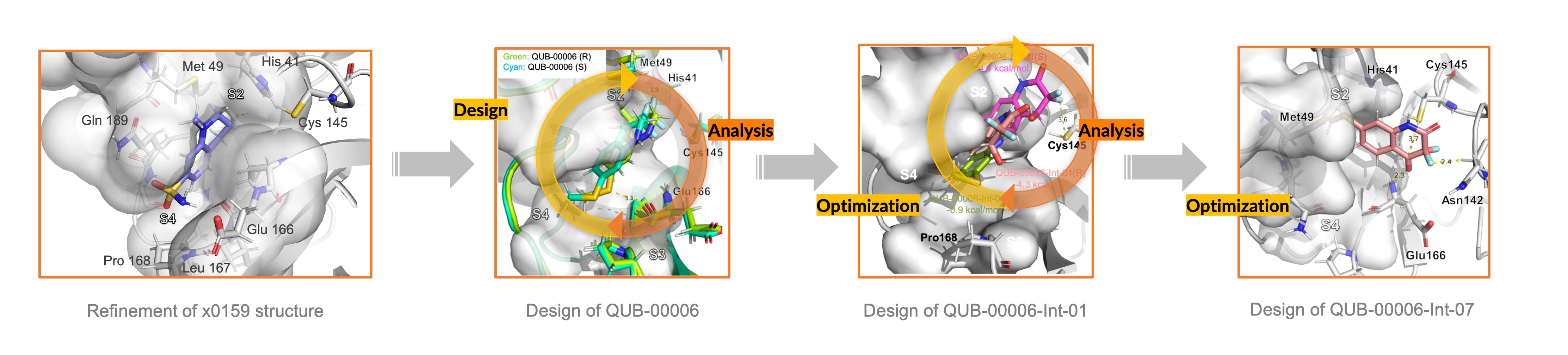}
\caption{{Schematic summary of our design strategy. First, we select a fragment with an available co-crystal structure with the protein. Then, to increase the binding to Mpro's subpockets, we design new-compounds by introducing chemical modifications to the known binder. Adaptive sampling simulations, free energy simulations, and NMR experiments were used to characterize the binding modes of the newly designed ligands. The promising binding modes were used as a starting point to design covalent ligands.}}
\end{figure*}
}

\section{Equilibration protocol prior to MD simulations}
Selected binding poses from the molecule docking were used for MD simulations.
The protein-ligand complexes were solvated in a cubic water box with a minimum distance between protein and box edge of 12 \si{\angstrom}. NaCl salt was added to neutralize the system and give a concentration of 0.15 mol/L.

The systems were equilibrated through multiple steps: (1) energy minimization to a threshold of 10 kcal/mol/\si{\angstrom^2}; (2) 0.1 ns NVT simulations at 200 K, 2.0 fs time step, with harmonic restraints (force constant 1.0 kcal/mol/\si{\angstrom^2}) on protein, ligand and crystallographic water; (3) 0.1 ns NVT simulations at 298 K, 2.0 fs time step, with harmonic restraints (force constant 1.0 kcal/mol/\si{\angstrom^2}) on protein and crystallographic water, and flat-bottom restraint (force constant 1.0 kcal/mol/\si{\angstrom^2}, radius 3.0 \si{\angstrom}) on the ligand. (4) 1 ns NPT simulations at 298 K and 1 bar, with same restraints as last step; (5) 6 ns NPT simulations at 298 K and 1 bar, with harmonic restraints (force constant 1.0 kcal/mol/\si{\angstrom^2}) on protein backbone.

Two independent equilibration simulations were conducted for each binding pose. The last 4 ns simulations for all binding poses of each ligand were used for clustering analysis. The ligand and heavy atoms of residues 41, 144, 145, 146, 163, 164 were used to calculate the RMSD. The DBSCAN algorithm with a cutoff distance of 0.8 \si{\angstrom} and the hierarchical clustering algorithm with 1.1 \si{\angstrom} cutoff for the average distance were used to select the largest clusters.

\section{SMILES of the studied compounds}
\begin{table}[h]
\begin{tabular}{|c|c|}
\hline
 Compound    & Smiles \\
 \hline
QUB-00006 & CSc1ccc2c(c1)NCC(F)(F)[C@@H]2C\\
\hline
QUB-00006-Int-07 & CSc1ccc2c(c1)NC(=O)C(F)(F)C2=O  \\
\hline
QUB-00006-Int-01 &   CSc1ccc2c(c1)NC(=O)C(F)(F)[C@@H]2O\\
\hline
x0195 & CN1CCCc2ccc(S(N)(=O)=O)cc21 \\
\hline
\end{tabular}
\caption{\label{Tab:deltasmiles}SMILES strings describing the studied compounds}
\end{table}



\section{Further analysis of QUB-00006-Int-01(R) adaptive sampling simulations}
We used the binary-version of nearest-neighbor heavy-atom distance contact between the ligand and MPro residues as feature vector.
In this version, if the nearest-neighbor heavy-atom distance between residues is smaller than a threshold, here set to 4 \si{\angstrom}, the contact is set 1, otherwise 0. This results in a contact binary feature vector associate with each trajectory frame, reducing the dimension to 608. The 4\si{\angstrom} threshold of the feature vector has been chosen in order to find the best compromise between locality and contact information. From then we performed another dimensionality reduction to retain only 10 slow collective variables (CVs): time-lagged variational autoencoder (TICA).TICA approximates the slow CVs by a linear combination of input coordinates. It has been shown in \citet{perez2013identification} that TICA finds the optimal approximation of slow CVs within the class of linear methods. For TICA, we chose a number of dimensions in order to cover $95\%$ of the kinetic variance and scaled them.
After optimization, we set the TICA lag-time to 3 ns and retained 12 dimensions. We then used the k-means clustering method on this reduced space. The k-means parameters were fixed to: 200 clusters, 200 maximum number of iteration and a tolerance factor of $10^{-12}$. Based on these clusters, we build a Hidden Markov State Model for which we tried different lag times. Here, we find converged values for lag times between 25ns and 50ns and we picked 40ns. After all these steps, we found 3 states to be relevant because of the relatively narrow conformational diversity and the not so pronounced time-scale gaps.
For each of the 3 clusters, we randomly extracted 100 structures. To analyze the clusters we focused on the residues which are in contact (within 4\si{\angstrom} threshold) with DS7(R) for > $20\%$ of the samples. We computed the mean and standard deviation of: 1) the heavy-atom distance between the ligand and a residue 2) residues RMSD and 3) total RMSD as is presented in the next figure. Although, such AS simulations cannot be considered as fully converged (see reference \citet{D1SC00145K} for a detailed discussion), they represent a powerful interpretative tools able to locate the potential binding modes. 
\begin{figure}[!htp]
\centering 
\subfloat{\label{a}\includegraphics[width=.5\linewidth]{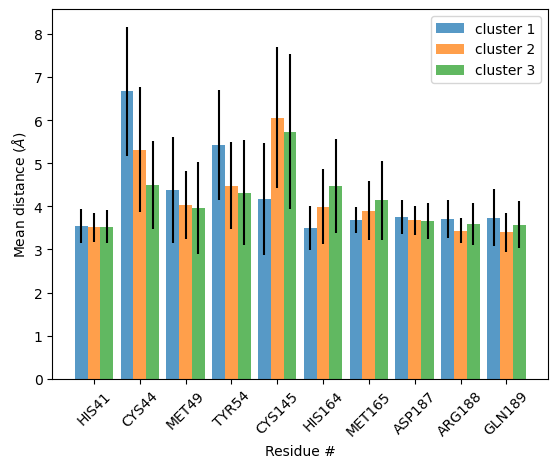}}\hfill
\subfloat{\label{b}\includegraphics[width=.5\linewidth]{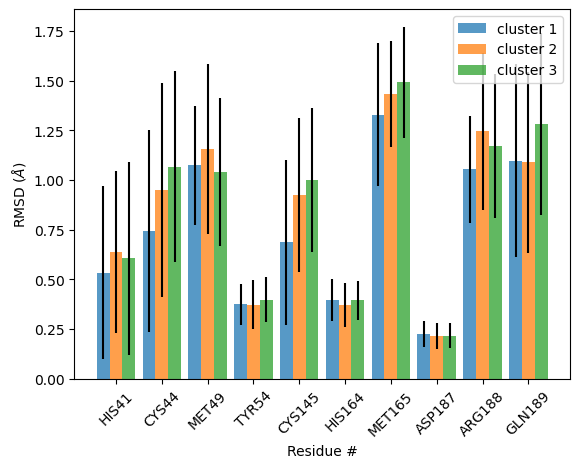}}\par
\subfloat{\label{c}\includegraphics[width=.5\linewidth]{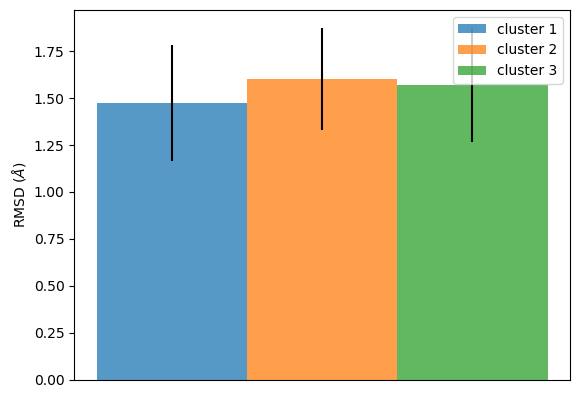}}
\caption{TICA clusters analysis: a) Heavy-atom distance distribution between drug and residues, b) residues RMSD distribution and c) total RMSD distribution.}
\end{figure}

\begin{figure}[!htp]
\centering 
\includegraphics[width=0.9\linewidth]{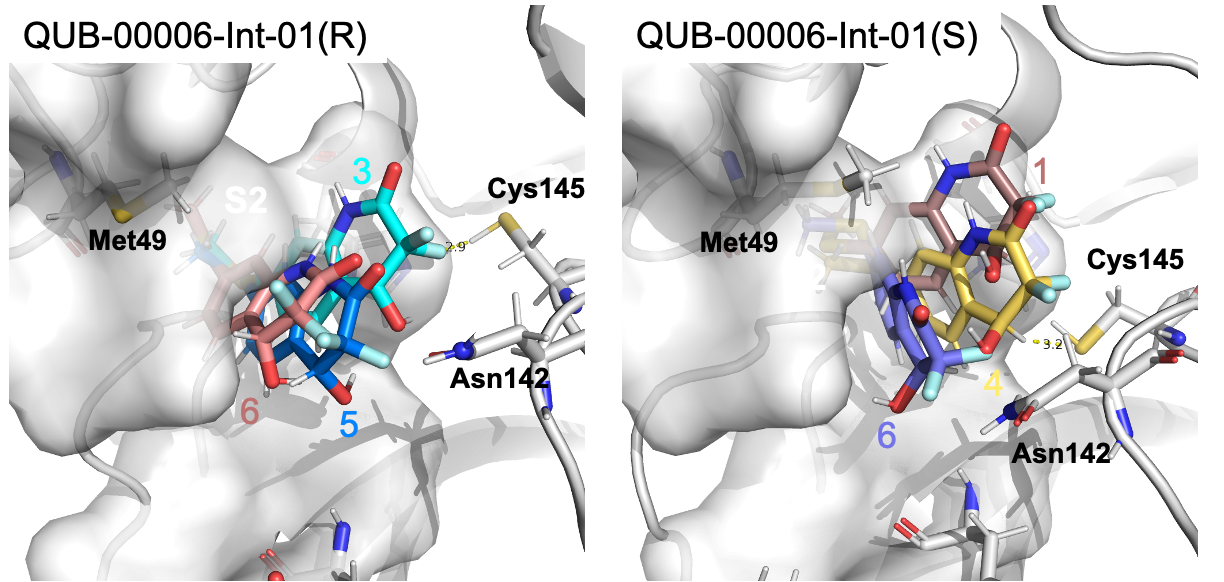}
\caption{Exploration of the binding mode of Qub-00006-Int-01 (R) and Qub-00006-Int-01 (S). A) Representative poses of the three clusters of Qub-00006-Int-01 (R) identified during MD simulations performed using our adaptive sampling approach. The 3 clusters represented by poses 3, 5, and 6 are consistent with the respective experimental binding mode of Qub-00006-Int-01 characterized by NMR (STD). Poses 3, 5, and 6 have close binding free energies since the calculative relative binding free energies show a maximum difference of 0.3 kcal/mol between the 3 pairs of clusters (Table 3). B) Representative poses of the three clusters of Qub-00006-Int-01 (S) identified during MD simulations performed using adaptive sampling. The 3 clusters represented by poses 1, 4, and 6 are consistent with the experimental binding mode of Qub-00006-Int-01 characterized by NMR(STD) and have close binding free energies with a difference of less than 0.2 kcal/mol between  the 3 pairs of clusters.
}
\end{figure}

\section{Experimental details}
\subsection{Compounds synthesis}
\textbf{Step 1} Synthesis of 2-amino-4-fluorobenzaldehyde 
A mixture of (2-amino-4-fluorophenyl)methanol (14.2 g, 100.61 mmol) and manganese(IV) oxide (52.48 g, 603.64 mmol) in THF (500 mL) was stirred several days (TLC control) at reflux. The solids were filtrated off and the filtrate was concentrated under reduced pressure to afford 2-amino-4-fluorobenzaldehyde (13.5 g, 97.03 mmol, 96.4\% yield).

\textbf{Step 2} Synthesis of 2-amino-4-(methylsulfanyl)benzaldehyde
To an ice-cooled solution of 2-amino-4-fluorobenzaldehyde (16.0 g, 115.0 mmol) in DMSO (50 mL) sodium methylsulfanide (16.12 g, 230.01 mmol, 76.77 ml, 2.0 equiv) was added dropwise. Upon completion of the reaction, the mixture was allowed to warm up to ambient temperature and stirred overnight. The solution was diluted with water and extracted with EtOAc(3x50mL). The organic layer was washed with water, brine, dried over sodium sulfate, and evaporated under reduced pressure to give 2-amino-4-(methylsulfanyl)benzaldehyde (13.0 g, 77.74 mmol, 67.6\% yield). 

\textbf{Step 3} Synthesis of 3,3-difluoro-4-hydroxy-7-(methylsulfanyl)-1,2,3,4-tetrahydroquinolin-2-one, QUB-00006$_$Int$_$01
To the mixture of activated zinc (3.6 g, 55.01 mmol) in dry THF, ethyl 2-bromo-2,2-difluoroacetate (9.71 g, 47.84 mmol) was added and the mixture was stirred for 1 hour at room temperature. Then, 2-amino-4-(methylsulfanyl)benzaldehyde (4.0 g, 23.92 mmol) in THF was added dropwise and the reaction mixture was stirred for a further 19 h at room temperature. The mixture was diluted with ethyl acetate (25x2 mL) and was washed with water (50x2 mL) and brine (50x2 mL), was dried over sodium sulfate, and concentrated under reduced pressure to give pure 3,3-difluoro-4-hydroxy-7-(methylsulfanyl)-1,2,3,4-tetrahydroquinolin-2-one, QUB-00006$_$Int$_$01. Yield: 1000 mg, 17\%; Appearance: Orange solid; 1H NMR (500 MHz, DMSO-d6) δ 10.93 (s, 1H), 7.34 (d, J = 8.1 Hz, 1H), 7.02 – 6.89 (m, 1H), 6.82 (s, 1H), 6.47 (d, J = 5.7 Hz, 1H), 5.00 – 4.84 (m, 1H), 2.44 (s, 3H); HPLC purity: 100\%; LCMS: 246.0[M+H]+.

\textbf{Step 4} Synthesis of 3,3-difluoro-7-(methylsulfanyl)-1,2,3,4-tetrahydroquinoline-2,4-dione, QUB-00006$_$Int$_$07.
A mixture of 3,3-difluoro-4-hydroxy-7-(methylsulfanyl)-1,2,3,4-tetrahydroquinolin-2-one (1.3 g, 5.3 mmol), manganese(IV) oxide (4.61 g, 53.0 mmol) and THF (50 mL) was stirred several days (TLC control) at reflux. The solids were filtrated out and the filtrate was concentrated to afford the 3,3-difluoro-7-(methylsulfanyl)-1,2,3,4-tetrahydroquinoline-2,4-dione, QUB-00006$_$Int$_$07. Yield: 1000 mg, 77.6\%; Appearance: Yellow solid; 1H NMR (400 MHz, DMSO-d6) δ 11.32 (s, 1H), 7.74 (d, J = 8.3 Hz, 1H), 7.06 (dd, J = 8.4, 1.8 Hz, 1H), 6.93 (d, J = 1.8 Hz, 1H), 2.54 (s, 3H); HPLC purity: 100\%; LCMS: 244.0[M+H]+.

\textbf{Step 5} Synthesis of 3,3-difluoro-4-hydroxy-4-methyl-7-(methylsulfanyl)-1,2,3,4-tetrahydroquinolin-2-one, QUB-00006$_$Int$_$09.
Chloro(methyl)magnesium (322.8 mg, 4.32 mmol, 1.47 ml, 3.0 equiv) was added dropwise to a solution of 3,3-difluoro-7-(methylsulfanyl)-1,2,3,4-tetrahydroquinoline-2,4-dione (350.0 mg, 1.44 mmol) in 30 mL of THF at -70 oC under argon and reaction mixture was stirred at this temperature 1 hour. Then it was quenched with 30mL of saturated aqueous NH4Cl solution and concentrated under vacuum. The residue was partitioned between 50 mL of water and 100 mL of EtOAc. The organic layer was washed with 20 mL of water, brine, dried over sodium sulfate, and concentrated under vacuum to give 3,3-difluoro-4-hydroxy-4-methyl-7-(methylsulfanyl)-1,2,3,4-tetrahydroquinolin-2-one, QUB-00006$_$Int$_$09. Yield: 350 mg, 86.3\%; Appearance: Yellow solid; 1H NMR (500 MHz, Chloroform-d) δ 8.58 (s, 1H), 7.47 (d, J = 8.2 Hz, 1H), 7.01 (dd, J = 8.1, 1.8 Hz, 1H), 6.75 (d, J = 1.9 Hz, 1H), 2.47 (s, 3H), 1.60 (s, 3H); HPLC purity: 100\%; LCMS: 260.2[M+H]+.

\textbf{Step 6} Synthesis of 3,3-difluoro-4-methyl-7-(methylsulfanyl)-1,2,3,4-tetrahydroquinolin-2-one, QUB-00006$_$Int$_$10
3,3-Difluoro-4-hydroxy-4-methyl-7-(methylsulfanyl)-1,2,3,4-tetrahydroquinolin-2-one (308.01 mg, 1.19 mmol) was dissolved in DCM, the mixture was cooled to 0 oC and 2,2,2-trifluoroacetic acid (677.28 mg, 5.94 mmol, 460.0 µl, 5.0 equiv) and triethylsilane (691.6 mg, 5.95 mmol, 950.0 µl, 5.0 equiv) was added. The mixture was heated to 40 oC and stirred 3h at that temperature.
After that mixture was cooled to ambient temperature and was washed with water and aq. NaHCO3 solution to give pure 3,3-difluoro-4-methyl-7-(methylsulfanyl)-1,2,3,4-tetrahydroquinolin-2-one, QUB-00006$_$Int$_$10. Yield: 200 mg, 69.2\%; Appearance: Beige solid; 1H NMR (600 MHz, DMSO-d6) δ 10.98 (s, 1H), 7.24 (d, J = 8.0 Hz, 1H), 6.96 (dd, J = 8.0, 1.9 Hz, 1H), 6.84 (d, J = 1.9 Hz, 1H), 3.71 – 3.50 (m, 1H), 2.43 (s, 3H), 1.24 (d, J = 7.1 Hz, 3H); HPLC purity: 100\%; LCMS: 244.0[M+H]+.

\textbf{Step 7}  Synthesis of 3,3-difluoro-4-methyl-7-(methylsulfanyl)-1,2,3,4-tetrahydroquinoline, QUB-00006$_$Int$_$08
To a solution of 3,3-difluoro-4-methyl-7-(methylsulfanyl)-1,2,3,4-tetrahydroquinolin-2-one (100.0 mg, 411.06 µmol) in 30 mL of dry THF borane dimethyl sulfide complex (93.73 mg, 1.23 mmol) was added in one portion. The resulting mixture was stirred at 45 oC overnight, then poured into cold K2CO3 aq. solution and extracted with EtOAc. The organic layer was washed with water, brine, dried over sodium sulfate, and evaporated under reduced pressure. The reside was subjected to HPLC to give 3,3-difluoro-4-methyl-7-(methylsulfanyl)-1,2,3,4-tetrahydroquinoline, QUB-00006$_$Int$_$08. Yield: 8.5 mg, 8.6\%; Appearance: Yellow oil; 1H NMR (400 MHz, DMSO-d6) δ 7.15 (d, J = 8.0 Hz, 1H), 6.75 (dd, J = 8.1, 2.0 Hz, 1H), 6.70 (d, J = 2.0 Hz, 1H), 3.76 – 3.62 (m, 1H), 3.62 – 3.53 (m, 1H), 3.39 – 3.23 (m, 1H), 2.55 (s, 3H), 1.51 (d, J = 7.0 Hz, 3H); HPLC purity: 100\%; LCMS: 230.2[M+H]+.

The synthesis of 3,3-difluoro-7-(methylsulfanyl)-1,2,3,4-tetrahydroquinolin-4-ol, QUB-00006$_$Int$_$02: to a solution of 3,3-difluoro-4-hydroxy-7-(methylsulfanyl)-1,2,3,4-tetrahydroquinolin-2-one (100.0 mg, 407.75 µmol) in 30 mL of dry THF borane dimethyl sulfide complex (92.93 mg, 1.22 mmol) was added in one portion. The resulting mixture was stirred at 45 oC overnight, then poured into cooled K2CO3 aqua solution and extracted with EtOAc. The organic layer was washed with water, brine, dried over sodium sulfate, and evaporated under reduced pressure. The resulting residue was subjected to HPLC purification to give 3,3-difluoro-7-(methylsulfanyl)-1,2,3,4-tetrahydroquinolin-4-ol, QUB-00006$_$Int$_$02. Yield: 48.9 mg, 51.9\%; Appearance: Beige solid; 1H NMR (400 MHz, Methanol-d4) δ 7.12 (d, J = 8.0 Hz, 1H), 6.57 (dd, J = 8.0, 1.8 Hz, 1H), 6.52 (d, J = 1.8 Hz, 1H), 4.56 (t, J = 7.6, 6.7 Hz, 1H), 3.69 – 3.51 (m, 1H), 3.42 – 3.34 (m, 1H), 2.42 (s, 3H); HPLC purity: 100\%; LCMS: 232.0[M+H]+.


\onecolumn

\begin{center}

\begin{longtable}{ccccc} 
\caption{\label{Tab:inactive} List of inactive compounds} \\ \hline
No. & ID & \makecell{MW\\ (g/mol)} & & Structure \\ \hline 
\endfirsthead

Table \ref{Tab:inactive} continued \\ \hline
No. & ID & \makecell{MW\\ (g/mol)} & & Structure \\ \hline 
\endhead

5  & Compound 6  & 232.25 &  & \includegraphics[height=10mm]{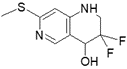 }  \\
8  & Compound 9  & 246.24 &  & \includegraphics[height=10mm]{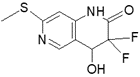 }  \\
9  & Compound 10 & 263.24 &  & \includegraphics[height=10mm]{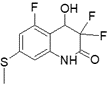}  \\
10 & Compound 11 & 231.27 &  & \includegraphics[height=10mm]{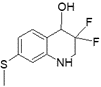}  \\
11 & Compound 13 & 264.25 &  & \includegraphics[height=10mm]{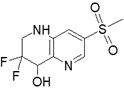}  \\
12 & Compound 14 & 248.25 &  & \includegraphics[height=10mm]{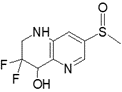}  \\
13 & Compound 15 & 184.62 &  & \includegraphics[height=10mm]{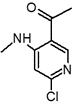}  \\
15 & Compound 19 & 190.03 &  & \includegraphics[height=10mm]{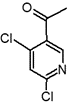}  \\
16 & Compound 20 & 263.26 &  & \includegraphics[height=10mm]{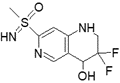}  \\
17 & Compound 21 & 323.37 &  & \includegraphics[height=10mm]{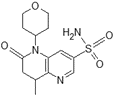}  \\
18 & Compound 23 & 196.27 &  & \includegraphics[height=10mm]{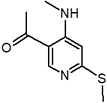}  \\
21 & Compound 26 & 232.25 &  & \includegraphics[height=10mm]{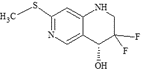}  \\
22 & Compound 27 & 232.25 &  & \includegraphics[height=10mm]{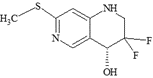}  \\
23 & Compound 28 & 206.26 &  & \includegraphics[height=10mm]{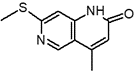}  \\
24 & Compound 29 & 243.27 &  & \includegraphics[height=10mm]{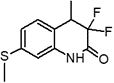}  \\
25 & Compound 30 & 259.27 &  & \includegraphics[height=10mm]{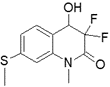}  \\
26 & Compound 31 & 259.27 &  & \includegraphics[height=10mm]{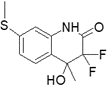}  \\
27 & Compound 32 & 231.26 &  & \includegraphics[height=10mm]{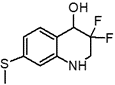}  \\
28 & Compound 33 & 246.23 &  & \includegraphics[height=10mm]{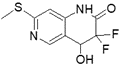}  \\
29 & Compound 34 & 277.24 &  & \includegraphics[height=10mm]{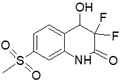}  \\
30 & Compound 35 & 274.29 &  & \includegraphics[height=10mm]{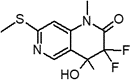}  \\
34 & Compound 37 & 217.17 &  & \includegraphics[height=10mm]{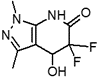}  \\
35 & Compound 38 & 274.24 &  & \includegraphics[height=10mm]{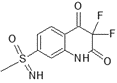}  \\
36 & Compound 39 & 257.26 &  & \includegraphics[height=10mm]{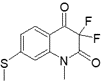}  \\ 
\hline

\end{longtable}

\end{center}
\
\twocolumn




\rmfamily 

\bibliography{rsc} 
\bibliographystyle{rsc}  